\documentclass[10pt]{article}
\usepackage{mathrsfs}
\usepackage{amsmath,amsfonts,amssymb,mathtools}
\usepackage{cite}
\usepackage{dsfont}
\usepackage{graphicx}
\usepackage{hyperref}
\usepackage{verbatim}
\usepackage{slashed}
\usepackage[all]{xy}
\usepackage{xcolor}
\usepackage{subfig}
\usepackage{tikz}
\usetikzlibrary{shapes.geometric,arrows}
\usepackage{relsize}
\usepackage{caption}
\usepackage{float}
\usetikzlibrary{positioning}
\usepackage{geometry}
\usepackage[T1]{fontenc}
\usepackage{lmodern}
\usepackage{stackengine} %per \ocirc 
 
\usepackage[utf8]{inputenc}

\hypersetup{
  colorlinks,
  citecolor=violet,
  linkcolor=blue,
  urlcolor=blue}

\csname @addtoreset\endcsname{equation}{section}
\DeclareMathAlphabet\mathbfcal{OMS}{cmsy}{b}{n}

\newcommand{\beq}{\begin{equation}}
\newcommand{\eeq}{\end{equation}}
\newcommand{\bea}{\begin{eqnarray}}
\newcommand{\eea}{\end{eqnarray}}
\newcommand{\ba}{\begin{array}}
\newcommand{\ea}{\end{array}}
\newcommand{\bit}{\begin{itemize}}
\newcommand{\eit}{\end{itemize}}
\newcommand{\nn}{\nonumber}

\newcommand{\mezzo}{\frac{1}{2}}
\newcommand{\complesso}{{\ \hbox{{\rm I}\kern-.6em\hbox{\bf C}}}}
\newcommand{\reale}{{\hbox{{\rm I}\kern-.2em\hbox{\rm R}}}}
\newcommand{\uno}{ \,  \raisebox{+0.14em}{{\hbox{{\rm \scriptsize ]}} \raisebox{-0.2em}{\kern-.8em\hbox{1}}}} \, }  %  operatore identit\`a

\renewcommand{\a}{\alpha}

\newcommand{\g}{\gamma}

\newcommand{\D}{\Delta}
\newcommand{\e}{\epsilon}
\newcommand{\Er}{{\mathbfcal{E}}}

\renewcommand{\k}{\kappa}

\renewcommand{\L}{\Lambda}

\newcommand{\m}{\mu}

\newcommand{\n}{\nu}
\renewcommand{\r}{\rho}
\newcommand{\s}{\sigma}

\renewcommand{\c}{\chi}

\newcommand{\x}{\xi}

\newcommand{\om}{\omega}

\begin{document}

%\begin{comment}

\begin{titlepage}

\vspace{0.3cm}

\begin{flushright}
%$IFUM$--1105--$FT$ \\
$LIFT$--5-3.23
\end{flushright}

\vspace{0.3cm}

\begin{center}
\renewcommand{\thefootnote}{\fnsymbol{footnote}}
%{\Huge \bf Accelerating, charged and rotating \\
%\vskip 5mm
%  NUT black holes - o -}
\vskip 9mm  
{\Huge \bf Equivalence principle and
\vskip 4mm
  generalised accelerating black holes
  \vskip 7mm
   from binary systems  }
\vskip 27mm
{\large {Marco Astorino$^{a}$\footnote{marco.astorino@gmail.com} %,
%Giovanni Boldi$^{b}$\footnote{giovanni.boldi@studenti.unimi.it}
}}\\

\renewcommand{\thefootnote}{\arabic{footnote}}
\setcounter{footnote}{0}
\vskip 8mm
\vspace{0.2 cm}
{\small \textit{$^{a}$Laboratorio Italiano di Fisica Teorica (LIFT),  \\
Via Archimede 20, I-20129 Milano, Italy}\\
} \vspace{0.2 cm}
%{\small \textit{$^{b}$Istituto Nazionale di Fisica Nucleare (INFN), Sezione di Milano \\
%Via Celoria 16, I-20133 Milano, Italy}\\
%} 
%\vspace{0.2 cm}

%{\small \textit{$^{b}$Universit\`a degli Studi di Milano}} \\
%{\small {\it Via Celoria 16, I-20133 Milano, Italy}\\
%}
\end{center}

\vspace{1.9 cm}

\begin{center}
{\bf Abstract}
\end{center}
{The Einstein equivalence principle in general relativity allows us to interpret accelerating black holes as a black hole immersed into the gravitational field of a larger companion black hole. Indeed it is demonstrated that C-metrics can be obtained as a limit of a binary system where one of the black holes grows indefinitely large, becoming a Rindler horizon. When the bigger black hole, before the limiting process, is of Schwarzschild type we recover usual accelerating black holes belonging to the Plebanski-Demianski class, thus type D. Whether the greater black hole carries some extra features, such as electric charges or rotations, we get generalised accelerating black holes which belong to a more general class, the type I. In that case the background has a richer structure, reminiscent of the physical features of the inflated companion, with respect to the standard Rindler spacetime.\\
This insight allows us to build a general type D metric, describing an accelerating Kerr-NUT black hole. It has well defined limits to all the type-D black holes of general relativity, including the elusive (type-D) accelerating Taub-NUT spacetime. Extension to the presence of the cosmological constant is also provided.}

\end{titlepage}

\addtocounter{page}{1}

\newpage

%\tableofcontents
%\newpage

\section{Introduction}
\label{sec:introduction}

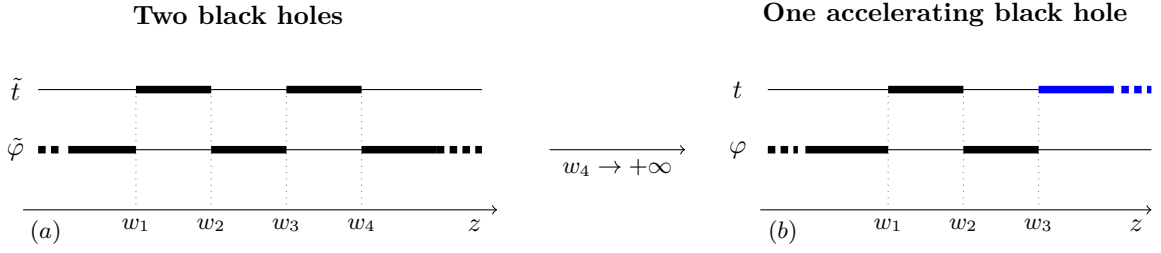
\begin{figure}[h!]
\centering
\begin{tikzpicture}

\draw[black,thin] (-10.3,-2) -- (-4.4,-2);
\draw[black,thin] (-9,-2.8) -- (-5,-2.8);
\draw[black,->] (-10.5,-3.6) -- (-4.2,-3.6);

\draw (-10.6,-2) node{$\tilde{t}$};
\draw (-10.6,-2.8) node{$\tilde{\varphi}$};
\draw (-4.5,-3.8) node{$z$};
\draw (-10.2,-3.9) node{{\small $(a)$}};
\draw (-9,-3.8) node{{\small $w_1$}};
\draw (-8,-3.8) node{{\small $w_2$}};
\draw (-7,-3.8) node{{\small $w_3$}};
\draw (-6,-3.8) node{{\small $w_4$}};
\draw (-7.7,-1) node{{{ \bf Two black holes}}};

\draw[gray,dotted] (-9,-2) -- (-9,-3.6);
\draw[gray,dotted] (-8,-2) -- (-8,-3.6);
\draw[gray,dotted] (-7,-2) -- (-7,-3.6);
\draw[gray,dotted] (-6,-2) -- (-6,-3.6);

\draw[black, dotted, line width=1mm] (-10.3,-2.8) -- (-10,-2.8);
\draw[black,line width=1mm] (-9.9,-2.8) -- (-9,-2.8);
\draw[black,line width=1mm] (-9,-2) -- (-8,-2);
\draw[black,line width=1mm] (-8,-2.8) -- (-7,-2.8);
\draw[black,line width=1mm] (-7,-2) -- (-6,-2);
\draw[black,line width=1mm] (-6,-2.8) -- (-5,-2.8);
\draw[black, dotted,line width=1mm] (-5,-2.8) -- (-4.4,-2.8);

\draw[black,->] (-3.5,-2.8) -- (-1.7,-2.8) node[midway, below, sloped] {{\small $w_4 \to +\infty$}};

\draw[black,thin] (-0.6,-2) -- (4,-2);
\draw[black,thin] (-0,-2.8) -- (4.5,-2.8);
\draw[black,->] (-0.7,-3.6) -- (4.5,-3.6);

\draw (-1,-2) node{$t$};
\draw (-1,-2.8) node{$\varphi$};
\draw (4.3,-3.8) node{$z$};

\draw[gray,dotted] (1,-2) -- (1,-3.6);
\draw[gray,dotted] (2,-2) -- (2,-3.6);
\draw[gray,dotted] (3,-2) -- (3,-3.6);

\draw (-0.4,-3.9) node{{\small $(b)$}};
\draw (1,-3.8) node{{\small $w_1$}};
\draw (2,-3.8) node{{\small $w_2$}};
\draw (3,-3.8) node{{\small $w_3$}};

\draw[black, dotted, line width=1mm] (-0.6,-2.8) -- (-0.2,-2.8);
\draw[black,line width=1mm] (-0.1,-2.8) -- (1,-2.8);
\draw[black,line width=1mm] (1,-2) -- (2,-2);
\draw[black,line width=1mm] (2,-2.8) -- (3,-2.8);
\draw[blue,line width=1mm] (3,-2) -- (4,-2);
\draw[blue, dotted,line width=1mm] (4.1,-2) -- (4.5,-2);

\draw (1.7,-1) node{{ {\bf One accelerating black hole}}};

\end{tikzpicture}
\caption{{\small $(a)$ Rod diagram for a collinear binary black hole system, associated to the Bach-Weyl metric. The $w_4 \to +\infty$ limit precisely gives the rod representation of the uncharged C-metric, or a single accelerating black hole $(b)$. Black segments on the time-like line represent event horizons while the semi-infinite blue line the Rindler horizon.}}
\label{fig:binary}
\end{figure}

Recently a new class of accelerating black holes has been discovered in general relativity. These solutions do not belong to the Plebanski-Demianski family \cite{Plebanski-Demianski}, thus are not of Type D in the Petrov classification \cite{Podolsky-nut}, \cite{PD-NUTs}, \cite{Type-I}. They are generalisations of the standard accelerating black holes, but they are endowed with extra features such as extra gravitomagnetic mass or a more general electromagnetic field. These novel features give rise to accelerating Taub-NUT black holes \cite{mann-stelea-chng}, \cite{Podolsky-nut} or accelerating Reissner-Nordstrom-NUT black holes \cite{tesi-giova}, \cite{PD-NUTs}, which were not known. Moreover the richer mathematical structure encoded in these type I black holes allows us to remove some of the unphysical properties of the standard metric, for instance it is possible to erase the Misner string from the accelerating Kerr-Newman-NUT spacetime, without completely renouncing to the NUT parameter nor introducing dramatic changes into the physical interpretation as a periodic time coordinate. Also it is possible to build accelerating Schwarzschild black holes in an electromagnetic background. Even though these solutions are built thanks to the Lie point symmetries of the Ernst equations, they are not much related with black hole embedded in external back-reacting Electromagnetic Bonnor-Melvin or rotational universes such as in the ones discovered in \cite{ernst-magnetic}, \cite{ernst-remove} or \cite{swirling}, \cite{marcoa-remove} respectively. \\
In \cite{Type-I} it has been proposed that these exotic charged accelerating black holes can be interpreted as a limit of a charged binary system where one of the black holes is enlarged indefinitely. We would like to consolidate this intuition and push it also for the standard Einstein theory. We would like to give a new interpretation to accelerating black holes, C-metrics and Plebanski-Demianski spacetimes, better explaining or motivating their typical features such as the origin of the acceleration or of their conical singularity. A new more general interpretation of these spacetime would open to the possibility of further extending their properties, hence constructing a more redefined physical model for accelerating black holes.\\
This new interpretation takes its first step from the well-known fact that the near (event) horizon limit of a Schwarzschild black hole is a Rindler horizon. In other words zooming close to the even horizon of a standard black hole, or making it very large, the event horizon becomes an accelerating horizon. What happens if in this picture we add a second black hole close by, thus we consider a binary system? If only one of the two gravitational sources grows bigger, then the model describes a small black hole in the proximity of a larger one.  \\
In order to answer this question more precisely, as an introductory exercise, we take the limit where one of the event horizons of the binary is enlarged indefinitely, so big that it reaches spatial infinity.   \\
The simplest metric which describes a binary black hole system is given, in vacuum general relativity,  by the Bach-Weyl metric \cite{bach-weyl}
\beq
\label{bach-weyl-metric}
          ds^2 = - \frac{\mu_1\mu_3}{\mu_2\mu_4} \ d\tilde{t}^2 + \frac{16\ \tilde{C}_f \ \mu_1^3 \mu_2^5 \mu_3^3 \mu_4^5 \ (d\r^2+dz^2)}{\mu_{12}^2 \mu_{14}^2 \mu_{23}^2 \mu_{34}^2 W_{13}^2 W_{24}^2 W_{11} W_{22} W_{33} W_{44}} + \rho^2 \frac{\mu_2 \mu_4}{\mu_1 \mu_3} \ d\tilde{\varphi}^2 \ , 
\eeq
where the solution is built, in cylindrical Weyl coordinates ($\r,z$) thanks to his basic solitonics ``bricks'' $\m_i(\r,z)$ defined as 
\begin{equation}
\label{mui}
\mu_i = w_i-z+\sqrt{\rho^2 + (z-w_i)^2} \ , \quad \qquad \mu_{ij} = (\mu_{i}-\mu_{j})^2  \ , \quad \qquad W_{ij} = \rho^2 + \mu_i\mu_j \ .
\end{equation}
In terms of the rod diagrams, which are an effective graphic representations of Weyl metrics \cite{emparan-reall}, the Bach-Weyl solution can be pictured as in figure (\ref{fig:binary}$.a)$. It represents two Schwarzschild black holes, where the two horizons are located at ($\r=0 , w_1<z<w_2$) and ($\r=0, w_3<z<w_4$). In the rod picture the event horizons correspond to the two black segments on the time-like line. Looking at picture we notice that the rod diagram of the binary system (\ref{fig:binary}$.a)$ becomes the one of an accelerating black hole (\ref{fig:binary}$.b)$ for $w_4 \to \infty$.

\begin{figure}[h]
%	\captionsetup[subfigure]{labelformat=empty}
	\centering
%	\hspace{-0.2cm}
	\includegraphics[scale=0.35]{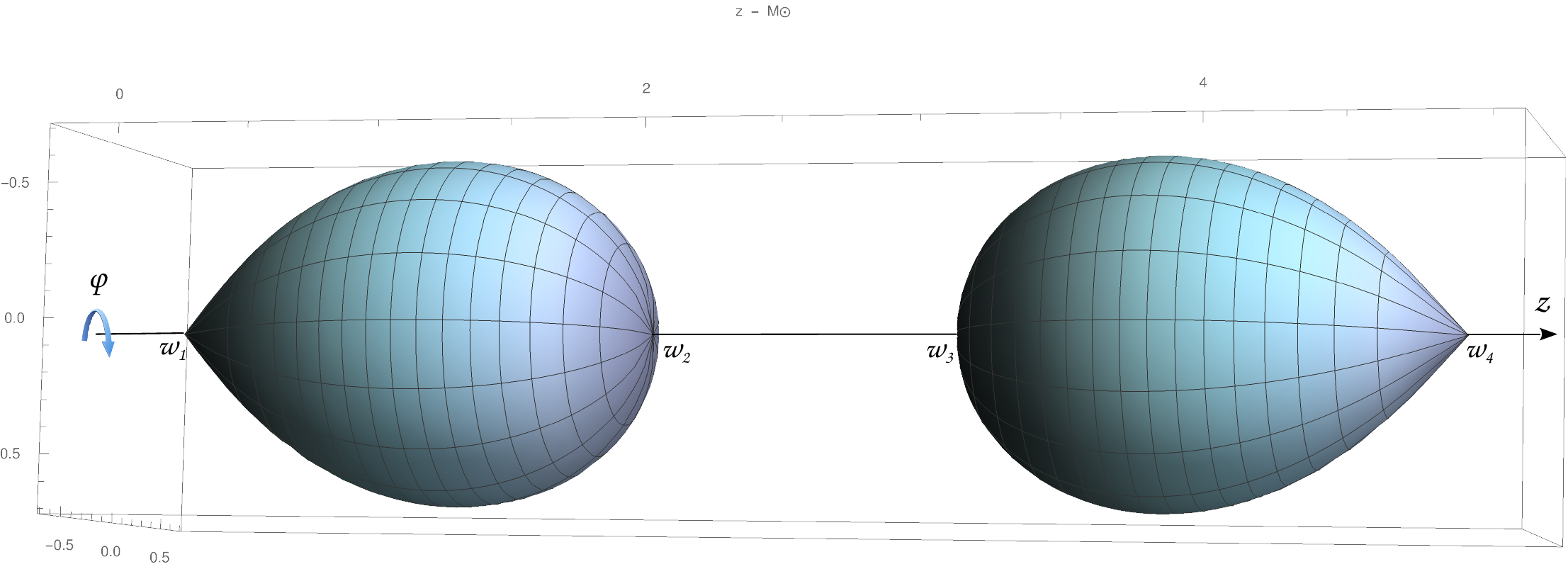}
	\caption{\small Embedding in the three dimensional Euclidean flat space of the event horizons of a Bach Weyl metric, which describes a couple of Schwarzschild black holes. In the pictures configuration the gauge freedom on $C_f$ is fixed to remove the conical singularity between the black holes, while $w_i=i$. Nevertheless conical singularities,  interpreted as semi-infinite strings, are present on the axis of symmetry, in the external region to balance an equilibrium configuration.}
	\label{fig:binario-sing}
\end{figure}

\begin{figure}[h!]
%	\captionsetup[subfigure]{labelformat=empty}
	\centering
%	\hspace{-0.2cm}
	\includegraphics[scale=0.26]{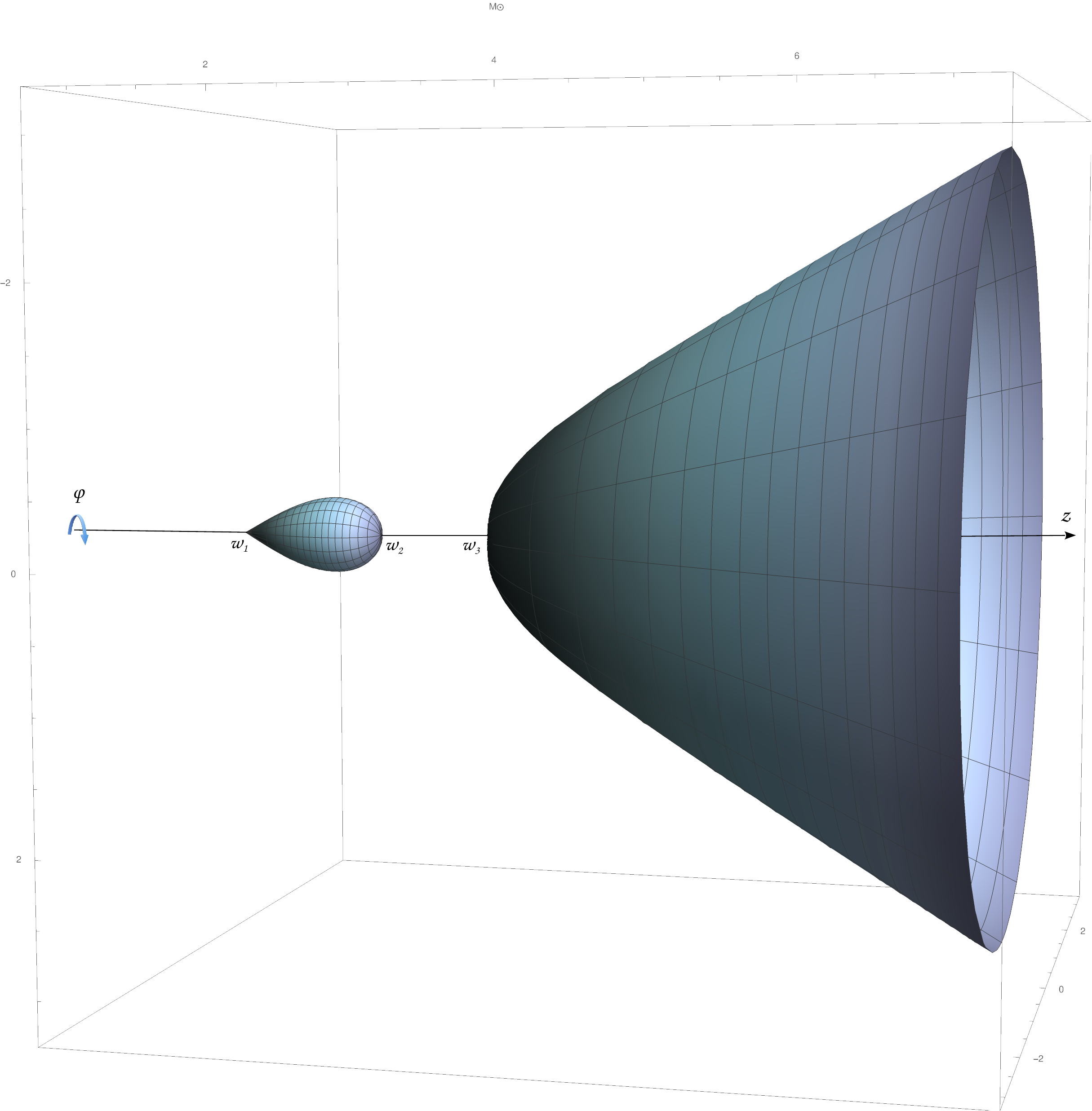}
	\caption{\small The $w_4 \to \infty$ limit brings a binary black hole system (illustrated in figure \ref{fig:binario-sing}) into an accelerating black hole metric (\ref{c-weyl}), whose embedding is pictured here. For larger values of $w_4$ the right element of the binary grows bigger and its event horizon becomes an accelerating horizon. If the right black hole of the binary carries no charges, then after the limit, the metric turns of special algebraic type, D, and the accelerating horizon is a usual Rindler one. When the right black hole of the binary carries some charges, the resulting accelerating metric remains of general type I, as the binary, and the accelerating background is endowed with that charge. The conical singularities on the symmetry axis eventually can be removed introducing external electromagnetic or gravitational fields, as done in \cite{ernst-remove} or \cite{many-rotating}.}
	\label{fig:c-sing}
\end{figure}

Because the correspondence between the rod representation and Weyl metrics, this observation suggests that also at the metric level this limit should hold. Pushing $w_4$ to spatial infinity means, in practice, to enlarge indefinitely the right black hole event horizon, while keeping the left black hole and the distance between the two sources finite.\\
This insight from the rods representation can be proven analytically. In fact just rescaling the time, the azimuthal angle and the $\tilde{C}_f$ such as
\beq
             \tilde{t} \to 2 w_4 t  \ , \hspace{1cm}  \tilde{\varphi} \to  \frac{\varphi}{2w_4} \ , \hspace{1cm} \tilde{C}_f \to (2w_4)^3 \ C_f   \ ,
\eeq 
and then taking the limit for $w_4 \to \infty$ we get the standard C-metric in Weyl coordinates
\beq \label{c-weyl}
           ds^2 = - \frac{\mu_1\mu_3}{\mu_2} \ dt^2 +  \frac{16\ C_f \ \mu_1^3 \mu_2^3 \mu_3^3  \ (d\r^2+dz^2)}{\mu_{12}^2 \mu_{23}^2 W_{13}^2 W_{11} W_{22} W_{33} } + \rho^2 \frac{\mu_2}{\mu_1 \mu_3} \ d\varphi^2 \ .
\eeq 
The accelerating Schwarzschild black hole metric in spherical-like coordinates ($t,r,x=\cos\theta,\varphi$) 
\beq \label{c-metric}
      ds^2 =   \frac{\displaystyle \left(1-\frac{2m}{r} \right) (\a^2r^2-1) dt^2 + \frac{dr^2}{ \left(1-\frac{2m}{r} \right) (1-\a^2r^2)} + \frac{r^2 \ dx^2}{(1+2m\a x)(1-x^2)} + r^2 (1+2m\a x) (1-x^2) \ d\varphi^2}{(1+\a rx)^2}
\eeq
can be retrieved thanks the diffeomorphism 
\bea
       \r &\to & \frac{\sqrt{(r^2-2mr)(1-\a^2 r^2)(1 + 2 m \a x)(1-x^2)}}{(1+\a rx)^2} \ \nn ,\\
       z &\to &  \frac{(\a r + x) [r-m(1-\a rx)]}{(1+\a rx)^2}  \ , \\
       t &\to & \sqrt{\a} t \ , \hspace{3.9cm}   \varphi \to \frac{\varphi}{\sqrt{\a}}  \ , \nn
\eea 
and choosing the free parameters in (\ref{c-weyl}) as follows 
\beq
         w_1 = -m \ ,  \hspace{1cm} w_2 = m \ , \hspace{1cm} w_3=\frac{1}{2\a} \ , \hspace{1cm}  C_f = \frac{m^2}{\a^3} \ .            
\eeq 
Therefore we have shown that if we enlarge indefinitely the event horizon of one of two components of the Bach-Weyl binary system (specifically the right one, by taking the limit $w_4 \to \infty$), while keeping the distance between the two hole finite and constant, we obtain the accelerating Schwarzschild black hole, or the basic C-metric. A similar conclusion was reached in \cite{Wang:1996sn}, with a more involved procedure. \\
Basically the event horizon of the blown black hole becomes a Rindler horizon. Thus we can interpret the C-metric as a Schwarzschild black hole near the horizon of a huge black companion. According to this interpretation, the reason why the black hole is accelerating is clearer: because it is immersed into the intense gravitational field generated by a huge and close black hole. Also this interpretation physically explains why it is not possible to regularise the accelerating black hole on both poles: since gravity is attractive, two sources alone tend to collapse one into the other; they can be keep apart only pulling them by some strings or inserting a strut to prevent their merging. According to this picture the cosmic string is not the source of the acceleration, but its tension maintains the small black hole at finite distance from the horizon of the big black hole.\\
Note that when the small black hole vanishes, this limit reduces to the well known fact that the near horizon limit of a Schwarzschild black hole is the Rindler horizon. On the other hand, according to the usual interpretation, where the Rindler horizon is caused by the acceleration of a string pulling the black hole, why the acceleration horizon remains when the black hole vanishes is less clear.  \\
A better graphical representation of the spacetimes involved in the limiting process can be achieved by drawing the surfaces of the horizons (before and after the $w_4\to\infty$ limit) embedded into the Euclidean flat space $\mathbb{E}^3$, as explained in \cite{Klemm-embedding} and pictured in figures \ref{fig:binario-sing} and \ref{fig:c-sing}.\\

\begin{figure}[H]
%	\captionsetup[subfigure]{labelformat=empty}
	\centering
%	\hspace{-0.2cm}
	\includegraphics[scale=0.31]{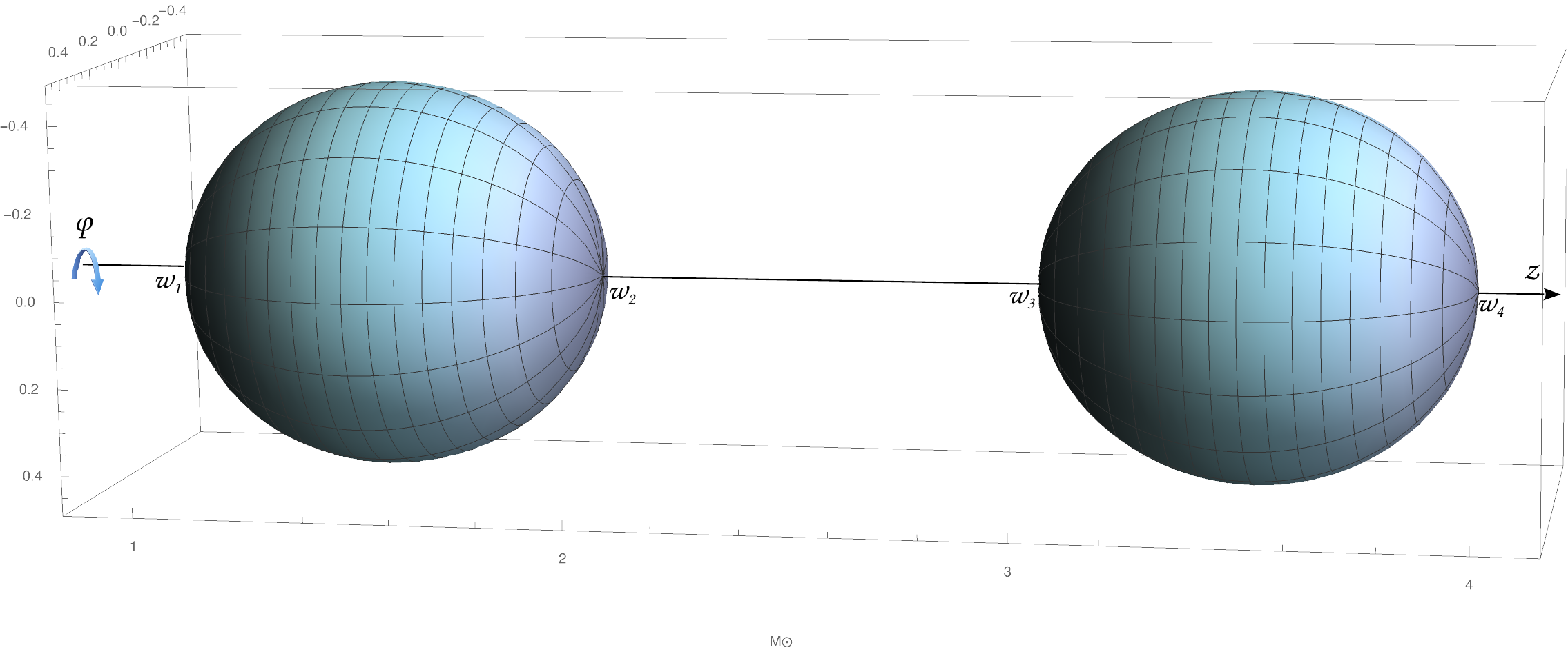}
	\caption{\small Embedding in $\mathbb{E}^3$ of the event horizons of the metric (\ref{reg-binary-metric})-(\ref{b2-bynary}). In the presence of an external gravitational field the binary black hole can be regularised and freed from conical singularities. In the picture $w_i=i,  C_f = 144, b_1 = 5/4 \log(4/3), b_2 = - \log(4/3)/4$ }
	\label{fig:binary-reg}
\end{figure}

\begin{figure}[H]
%	\captionsetup[subfigure]{labelformat=empty}
	\centering
	\includegraphics[scale=0.42]{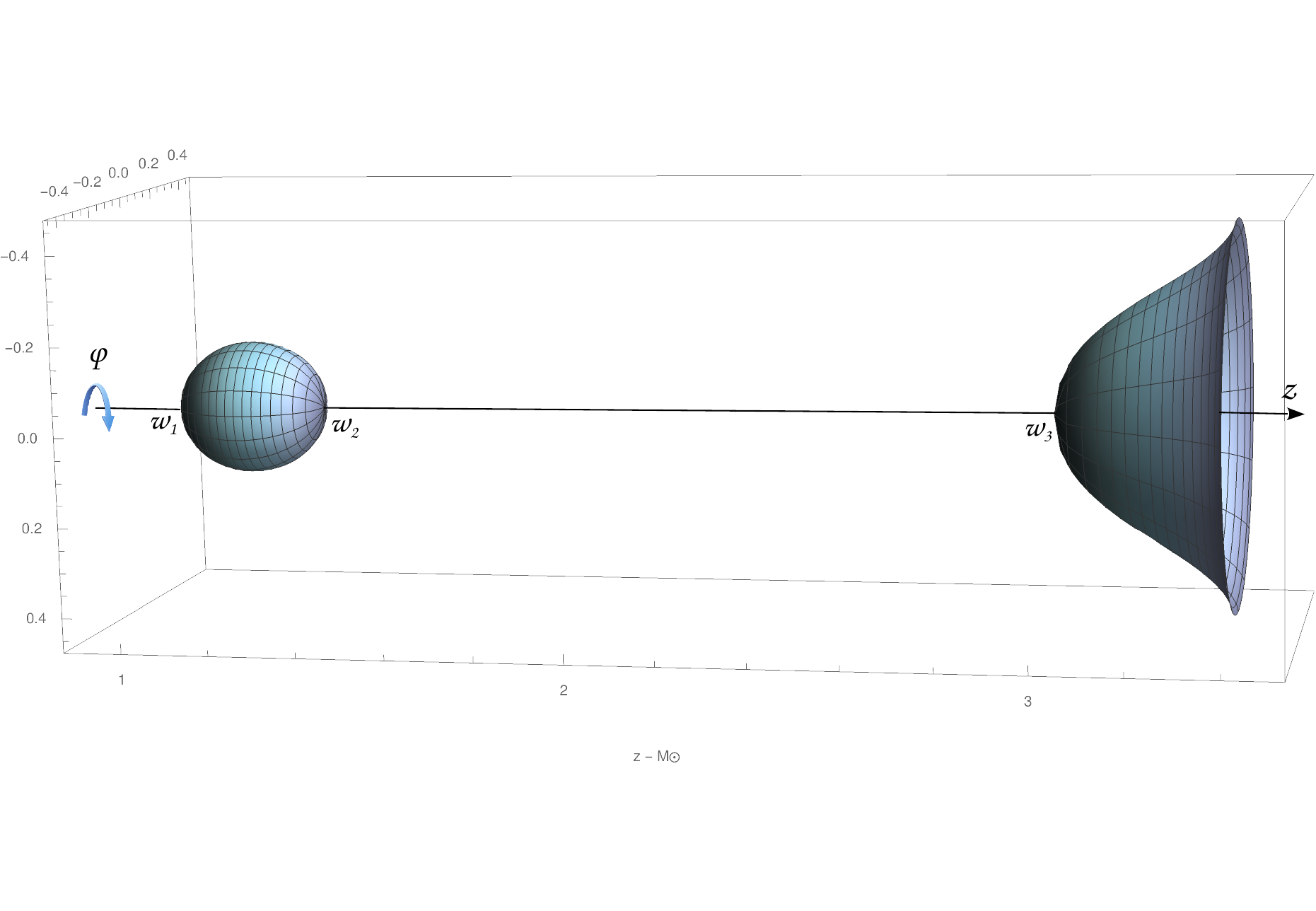}
	\caption{\small  The limit for $w_4 \to \infty$ can be carried on also for the regularised binary metric (\ref{reg-binary-metric})-(\ref{b2-bynary}), pictured in figure \ref{fig:binary-reg}. The embedding of the horizons of the resulting metric, described by (\ref{reg-c-metric})-(\ref{b1-reg-cmetric}), are pictured here, $b_2 = -1/15$.}
	\label{fig:acc-bh-reg}
\end{figure}

As can be appreciated also from the graphics both the binary and the single accelerating spacetime are affected by conical singularities. A delta-like matter-energy axial distribution that is needed to maintain an equilibrium configuration between the sources or in the external region avoids the gravitational collapse. It can be interpreted as a couple of semi-infinite strings pulling one horizon along the z-axis toward infinity or alternatively as a finite length strut (of repulsive matter, which violates all the reasonable energy conditions) that support and prevent the merging of the two horizons. A gauge freedom on the parameter $C_f$ can be used to reabsorb at most one of these excess or deficit angle of the metric, but generically these possible conicities are three for the binary system or two for the C-metric. In fact conical singularities may be encountered on the axis of symmetry, in the regions between the horizons. Hence, as can be seen in the pictures, we can have three regions for the double black hole (i.e. $z < w_1, \ w_2 < z < w_3, \ z>w_4$) and two in the case of the single black hole ($z < w_1, \ w_2 < z < w_3$).  Since these singularities are problematic both from a theoretical and phenomenological point of view, mechanisms for complete removal of these defects have been studied in the literature, basically by adding back-reacting gravitational or electromagnetic backgrounds. The interaction of a charged black hole with the external Maxwell field \cite{ernst-remove}, the presence of external gravitational sources \cite{ernst-generalized-c}, or the gravitational spin-spin interaction between the angular momentum of a black hole and the rotating background frame dragging \cite{marcoa-remove} are known methods to regularise the spacetime and, at the same time, they provide a plausible physical reason for the acceleration. \\
In \cite{many-rotating} it has been shown how to add a generic multipolar expansion to any axisymmetric and stationary spacetime in general relativity. We can exploit that result to regularise both the binary system and the accelerating metric and remove everywhere the conical singularities embedding our metric into a suitable multipolar background. For sake of simplicity we can focus just on the first two terms in the external gravitational multipolar expansion, which are sufficient to guarantee a non-conical geometry without constraining the characteristic physical parameters of the black holes, such as the masses or the distances between the sources. In practice, as a starting point, instead of the Bach-Weyl metric (\ref{bach-weyl-metric}), we consider the binary system at equilibrium generated in \cite{marcoa-binary}:
\beq \label{reg-binary-metric}
       ds^2 = - \frac{\mu_1\mu_3}{\mu_2\mu_4} e^{2b_1z+2b_2z^2-b_2\r^2} \ d\tilde{t}^2 + \frac{16\ \tilde{C}_f \ \mu_1^3 \mu_2^5 \mu_3^3 \mu_4^5 \ \tilde{k}(\r,z) \ (d\r^2+dz^2)}{\mu_{12}^2 \mu_{14}^2 \mu_{23}^2 \mu_{34}^2 W_{13}^2 W_{24}^2 W_{11} W_{22} W_{33} W_{44}} + \rho^2 \frac{\mu_2 \mu_4}{\mu_1 \mu_3} e^{-2b_1z-2b_2z^2+b_2\r^2} \ d\tilde{\varphi}^2 , 
\eeq
where
\beq
     \tilde{k}(\r,z) := e^{-b_1^2\r^2-2b_1(z+2b_2z\r^2-\m_1+\m_2-\m_3+\m_4) + b_2 [\frac{b_2}{2}\r^4-2z^2(1+2b_2\r^2)+4w_1\m_1-\m_1^2-4w_2\m_2+\m_2^2+4w_3\m_3-\m_3^2-4w_4\m_4+\m_4^2]} \ .\nn
\eeq
It is a generalisation of the singular binary (\ref{bach-weyl-metric}), which is recovered for $b_1=0$ and $b_2=0$. The regularity on the axis of symmetry is enforced, on the two external regions ($z<w_1$, $z>w_4$) and on the central one ($w_2<z<w_3$), by fixing the gauge constant $C_f$ and the external field parameters $b_i$ such that
\bea 
       C_f &=& 16 (w_2-w_1)^2 (w_3-w_2)^2 (w_4-w_1)^2 (w_4-w_3)^2 \\
       b_1 &=&  - \frac{(w_4^2-w_3^2+w_2^2-w_1^2) \log\left[ \frac{(w_4-w_1)(w_3-w_2)}{(w_3-w_1)(w_4-w_2)} \right]}{2(w_2-w_1)(w_4-w_3)(w_4+w_3-w_2-w_1)}  \\
       b_2 &=&  \frac{(w_4-w_3+w_2-w_1) \log\left[ \frac{(w_4-w_1)(w_3-w_2)}{(w_3-w_1)(w_4-w_2)} \right]}{2(w_2-w_1)(w_4-w_3)(w_4+w_3-w_2-w_1)} \label{b2-bynary}
\eea
Following the limiting procedure illustrated above, where one of the black hole is enlarged infinitely we get an accelerating black hole but without any conical singularity. 
\beq \label{reg-c-metric}
           ds^2 = - \frac{\mu_1\mu_3}{\mu_2} e^{2b_1z+2b_2z^2-b_2\r^2} \ dt^2 +  \frac{16\ C_f \ \mu_1^3 \mu_2^3 \mu_3^3\ k(\r,z)  \ (d\r^2+dz^2)}{\mu_{12}^2 \mu_{23}^2 W_{13}^2 W_{11} W_{22} W_{33} } +  \frac{\rho^2 \mu_2}{\mu_1 \mu_3} e^{-2b_1z-2b_2z^2+b_2\r^2} \ d\varphi^2 \ ,
\eeq 
with 
\beq
            k (\r,z) = e^{-b_1^2\r^2+2b_1(z-2b_2z\r^2+\m_1-\m_2+\m_3) + \frac{b_2}{2}[b_2\r^4+z^2(4-8b_2\r^2)+2(\r^2+4w_1-\m_1^2-4w_2\m_2+\m_2^2+4w_3\m_3-\m_3^2)]} \ .
\eeq
The constraints on $C_f$ and $b_1$ that assure the absence of axial singularities become
\bea
         C_f &=& e^{4 b_2 ( w_1 + w_2 - w_3) w_3} (w_2-w_1)^2 (w_3-w_1)^2 \left(\frac{w_3-w_2}{w_3-w_1} \right)^{\frac{2w_3}{w_2-w_3}} \ , \\
         b_1 &=& -b_2 (w_1+w_2) - \frac{\log \left(\frac{w_3-w_2}{w_3-w_1} \right)}{2(w_2-w_1)} \ . \label{b1-reg-cmetric}
\eea
As can be seen from the pictures \ref{fig:binary-reg} and \ref{fig:acc-bh-reg} of the embedded horizons into $\mathbb{E}^3$, the surfaces for both the binary and the accelerating single black hole now appear smooth. That's because the external multipolar expansion sustains the equilibrium configurations.
Note that this external gravitational fields model actual sources at large spatial distances, such as galaxies \cite{deCastro}, therefore the background metric presents curvature singularity at spatial infinity. This is in line with the matter interpretation of the external gravitational field however it would be better associated with a matter distribution to dress these divergences, otherwise the model, in particular for the binary system, is best suited for only a local description of the black hole couple. The accelerating black hole picture may suffers, to a lesser extent, from this issue because of the screen provided by the Rindler horizon, which cover the curvature singularities at large space-like distances.\\

\section{Type-I NUTty C-metrics from binary black holes system}

A natural question now arises: if we include in the above limiting procedure the NUT parameter, would we be able to obtain an accelerating black hole endowed also with gravitomagnetic mass? More specifically if we consider a static binary system with NUT charge and we enlarge one of the two event horizons, we will obtain a NUTty C-metric? This question is non trivial because NUTty accelerating black holes, without angular momentum, has been unknown for a long time. In fact C-metrics were though to be incompatible with the NUT parameter because these kinds of spacetimes do not belong to the Plebanski-Demianski family, therefore they are outside the D type of the Petrov classification. However that solution was found in \cite{mann-stelea-chng} and recently it has been put in a convenient parametrisation in \cite{Podolsky-nut}. Then an efficient way to build a large family of gravitomagnetic and accelerating spacetime including the Kerr-Newman family was established in \cite{PD-NUTs}, see also \cite{Type-I} for generalisations. This method relies on the Lie-point symmetries of the complex Ernst equations
\beq
     \label{ernst-eq}  \big( \textsf{Re} \ \Er \big) \nabla^2 \Er  =   \overrightarrow{\nabla} \Er  \cdot \overrightarrow{\nabla} \Er   \ ,       \\
\eeq
where the complex gravitational Ernst potential $ \Er(\r,z) := f(\r,z)  + i \ h(\r,z) $, stems from the metric (\ref{LWP-metric}) \cite{ernst2}
and\footnote{The vector differential operators are the usual standard gradient or Laplacian in flat cylindrical coordinates, while  the vectorial base is  ($\overrightarrow{e}_\r , \overrightarrow{e}_\varphi , \overrightarrow{e}_z$), more details can be found in \cite{enhanced}.} 
\beq
       \label{h-e}    \overrightarrow{\nabla} h := - \frac{f^2}{\r} \overrightarrow{e}_\varphi \times \overrightarrow{\nabla} \omega  \ .
\eeq 
Ernst equations, indeed, are equivalent to the vacuum Einstein equations, $R_{\m\n}=0$ for axisymmetric and stationary spacetimes described by the Lewis-Weyl-Papapetrou metric
\beq
      \label{LWP-metric}
{ds}^2 = -f ( dt - \omega d\varphi)^2 + f^{-1} \bigl[ e^{2\gamma}  \bigl( {d \rho}^2 + {d z}^2 \bigr) +\rho^2 d\varphi^2 \bigr] \ \   .
\eeq
The vacuum Ernst equations enjoys a SU(1,1) group of Lie-point symmetries, however the only continuos transformation with a non trivial action on the metric is the Ehlers transformation
\beq \label{ehlers}
\Er \longrightarrow \bar{\Er} = \frac{\Er}{1 + i c \Er} \ .
\eeq
This one parameter transformation leaves invariant the Ernst equations (\ref{ernst-eq}), hence when acting on a massive solution of the theory it maps electrovacuum solutions in electrovacuum solution, but changing the physical properties of the metrics. Specifically it rotates the gravitational mass into the gravitomagnetic mass, see section 4.1 of \cite{PD-NUTs} for details. In practice, depending on the value of the real rotating parameter $c$, it can add to a massive solution the NUT ``charge''.\\
When we apply the Ehlers transformation (\ref{ehlers}) to the Bach-Weyl solution we obtain\footnote{For computational details see \cite{enhanced}, where an electrically charged double black hole with gravitomagntetic mass was built.} the Bach-Weyl-NUT metric in the form of (\ref{LWP-metric}) with 
\bea \label{Bach-Weyl-NUT}
         f(\r,z) &=& \frac{\m_1 \m_2 \m_3 \m_4}{c^2\m_1^2\m_3^2 + \m_2\m_4} \ ,  \\
         \omega(\r,z) &=& 2 c \ (\m_1-w_1-\m_2+w_2+\m_3-w_3-\m_4+w_4) + \omega_0 \ , \nn \\
         e^{2\g(r,x)}    &=&    \frac{16\ \tilde{C}_f \ \mu_1^4 \mu_2^4 \mu_3^4 \mu_4^4}{\mu_{12}^2 \mu_{14}^2 \mu_{23}^2 \mu_{34}^2 W_{13}^2 W_{24}^2 W_{11} W_{22} W_{33} W_{44}}        \ . \nn
\eea
This spacetime represents a couple of Taub-NUT black holes, each with its own independent mass but with NUT parameters which are not independent, because (apart their distance) they have a functional dependence on just 3 physical parameters: $c$ and the two masses. It easy to check that interpretation for the metric, just by sending the first or alternatively the second couple of the four solitons, which constitute the solution, to plus or minus infinity respectively. For instance when $w_3$ and $w_4$ goes to infinity simultaneously we remain precisely with the Taub-NUT metric. \\

On the other hand taking exactly the same limit\footnote{Taking care of rescaling also the NUT parameter as $c \to w_4 c$.} of the section above for the basic Bach-Weyl solution, where the right black hole grows indefinitely while keeping finite and constant the distance with the left companion, we get  
\beq
ds^2 = - \frac{4 \m_1 \m_2 \m_3}{4\m_2+c^2\m_1^2\m_3^2} [dt-c(2z+\m_1-\m_2+\m_3+w_0)d\varphi]^2 + \frac{4\m_2+c^2\m_1^2\m_3^2}{4 \m_1 \m_2 \m_3} \left[ \frac{16 C_f \m_1^4 \m_2^2 \m_3^4 (d\r^2+dz^2)}{\m_{12}^2 \m_{23}^2 W_{11} W_{22} W_{33}W_{13}^2} + \r^2 d\varphi^2 \right] .
\eeq
This metric represents an accelerating black hole endowed with gravitomagnetic mass, first found in \cite{mann-stelea-chng}. By the following change of coordinates
\bea
            \r &=& \frac{\sqrt{[1-\a^2(r - \tilde{r}_+)^2 ](r-\tilde{r}_+)(r-\tilde{r}_-)[1-\a(\tilde{r}_+-\tilde{r}_-)](1-x^2)}}{[1+\a(r-\tilde{r}_-)]^2} \\
            z  &=& \frac{[x + \a(r - \tilde{r}_+)]\{r-\tilde{r}_- - \mezzo(\tilde{r}_+-\tilde{r}_-)[1-\a x (r-\tilde{r}_-)]\}}{[1+\a(r-\tilde{r}_-)]^2}
\eea
and choosing the parameters such as 
\beq
         C_f = \frac{m^2}{\a^3} \ ,  \hspace{1 cm} w_1 = - \frac{1}{2} \left( \tilde{r}_+ - \tilde{r}_- \right) \ , \hspace{1cm} w_2 =  \frac{1}{2} \left( \tilde{r}_+ - \tilde{r}_- \right) \ , \hspace{1cm}  w_3 = \frac{1}{2\a} \ ,
\eeq
where 
\beq
        \tilde{r}_\pm = m \pm \sqrt{m^2 + \ell^2} \ , \hspace{1cm} c = \sqrt{-\frac{\tilde{r}_-}{\tilde{r}_+}}
\eeq
we can write the metric above, as in \cite{Podolsky-nut}, in the following form
\beq
 ds^2 = -f(r,x) \left[ dt - \omega(r,x) d\varphi \right]^2 + \frac{1}{f(r,x)} \left[ e^{2\gamma(r,x)}  \left( \frac{{d r}^2}{\Delta_r(r)} + \frac{{d x}^2}{\Delta_x(x)} \right) + \rho^2(r,x) d\varphi^2 \right] \ \ ,
\eeq 
with 
\bea
         f(r,x) &=&      \frac{-(\tilde{r}_+-\tilde{r}_-) \D_r(r)\Omega^2(r,x)}{\tilde{r}_-\left[r-\tilde{r}_+ -\frac{(r-\tilde{r}_-)\a^2 \D_r(r)}{1-\a^2(r-\tilde{r}_-)} \right]^2 + (r-\tilde{r}_-)^2\tilde{r}_+\Omega^4(r,x)}     \ ,  \nn  \\  
         \omega(r,x) &=&  - \frac{2 \sqrt{-\tilde{r}_+ \tilde{r}_-} \{1 + \a(r - \tilde{r}_-)[\a (r-\tilde{r}_-)-2x)] \} \D_x(x) }{(\tilde{r}_+-\tilde{r}_-)(1-x^2)\a \Omega^2(r,x)} + \tilde{\omega}_0         \ ,  \nn  \\
         \g(r,x) &=& \mezzo \log \left[ \frac{\D_r(r)}{\Omega^4(r,x)} \right] \ , \nn  \\         
         \r(r,x) &=& \frac{\sqrt{\D_r(r) \D_x(x)}}{\Omega^2(r,x)} \ , \nn \\
         \Omega(r,x) &=&  1 - \a x (r-\tilde{r}_-)   \ ,   \nn  \\     
         \D_r(r)   &=& [(r-\tilde{r}_+)(r-\tilde{r}_-)][1-\a^2(r-\tilde{r}_-)^2] \ , \nn \\
                     \D_x(x) &=&   (1-x^2)[1-\a x(\tilde{r}_+-\tilde{r}_-)]  \ .
\eea

It is easy to see that the above solution for zero NUT parameter: $\ell=0$ (so $ c=0, \ \tilde{r}_-=0, \ \tilde{r}_+=2m$), reduces to the standard C-metric (\ref{c-metric}), while for zero acceleration parameter, i.e. $\a=0$, it recovers the Taub-NUT spacetime. Charged and rotating generalisations have been achieved in \cite{PD-NUTs}, further extensions can be found in \cite{Type-I}.
  
Hence the limit of the double Taub-NUT black hole (\ref{Bach-Weyl-NUT}), for one of the event horizon growing indefinitely, gives the accelerating type-I Schwarzschild-NUT metric, or equivalently an accelerating Taub-NUT black hole near the event horizon of another, but huge, Taub-NUT companion.  \\

\section{Generalised accelerating black holes from binary systems}
\label{binaryD}

The most generic double black hole system that can be written from the inverse scattering technique in general relativity \cite{belinski-book}, describes a couple of Kerr-NUT black holes aligned on the axis of symmetry and at a finite proper distance. For $n$ even, the general metric that describes $n/2$ axially aligned, stationary rotating and axisymmetric black hole can be expressed as
\beq \label{metric-ist}
          ds^2 = g_{ab}(\rho,z) dx^a dx^b + f(\r,z)(d\r^2+dz^2)
\eeq
with $a,b,c \in \{0,1\} $ and $k,l \in \{1,..., n\}$, so $x^a = \{t,\varphi\}$. The metric components take the form
\begin{subequations}
\label{metric-ph}
\begin{align}
\label{g-ph}
g_{ab}(\rho,z) & =  \frac{1}{\rho^n} \Biggl(\prod_{k=1}^n \mu_k\Biggr) \Biggl[ \ \overset{\circ}{g}_{ab} - \sum_{k,l=1}^n \frac{(\Gamma^{-1})_{kl} L_a^{(k)} L_b^{(l)}}{\mu_k\mu_l}\Biggr] \, , \\
\label{f-ph}
f(\rho,z) & =  \frac{16 \ C_f \overset{\circ}{f}_0}{\rho^{n^2/2}} \ \Biggl( \prod_{k=1}^n \mu_k \Biggr)^{n+1} \ \Biggl[ \prod_{k>l=1}^n (\mu_k-\mu_l)^{-2} \Biggr] \det\Gamma \, ,
\end{align}
\end{subequations}
where $L^{(k)}_a = m^{(k)}_c \overset{\circ}{g}_{ca}$. The $(\rho,z)$ sector of the background Minkowski metric in cylindrical Weyl coordinates is described by  $\overset{\circ}{f}=1$, while the $(t,\varphi)$ part is encoded in $\overset{\circ}{g}$ as follows
\beq
        \overset{\circ}{g}_{ab} = \bigg( \begin{tabular}{cc}
        -1 & 0 \\ 
        0 & $\rho^2$ 
        \end{tabular} \bigg)    \hspace{3mm} , \hspace{12mm} \Gamma_{kl} = \frac{m_a^{(k)} \  \overset{\circ}{g}_{ab} \ m_b^{(l)}}{\rho^2 + \mu_k \mu_l}\hspace{3mm} , \hspace{12mm}  m^{(k)}_a = \left( C_0^{(k)} , \frac{ C_1^{(k)}}{\m_k} \right) \ ,
\eeq

$n$ solitons bring in the metric $2n$ physical integration constants  
\begin{subequations}\label{C01}
\begin{align}
C_1^{(2i-1)}C_0^{(2i)} - C_0^{(2i-1)}C_1^{(2i)} & = \sigma_i \ \, , \qquad
C_1^{(2i-1)}C_0^{(2i)} + C_0^{(2i-1)}C_1^{(2i)} = -m_i \, , \\
C_0^{(2i-1)}C_0^{(2i)} - C_1^{(2i-1)}C_1^{(2i)} & = \ell_i \ , \qquad
C_0^{(2i-1)}C_0^{(2i)} + C_1^{(2i-1)}C_1^{(2i)} = a_i \ .
\end{align}
\end{subequations}
Here $m_i$, $a_i$, $\ell_i$ are respectively related to the mass, angular momentum and the NUT parameters, with $i\in{1,n/2}$.
We have also defined
$\sigma_i^2\equiv m_i^2 - a_i^2 + \ell_i^2$.
The ordered poles $w_k$, with $w_k < w_{k+1}$, are taken as follows 
\beq \label{wi}
w_1 = z_1 - \sigma_1 \ , \quad
w_2 = z_1 + \sigma_1 \ , \quad ... \qquad 
w_{2i-1} = z_i - \sigma_i \ , \quad
w_{2i} = z_i + \sigma_i \ ,
\eeq
where $z_i$ denotes the position of the $i-$black hole. In this case because the symmetry along the $z$-axis, only the distance between the centers of the sources, $|z_{i+1}-z_{i}|$ is relevant.

When $n=4$, four gravitational solitons, defined by (\ref{mui}), are added to the background metric. Since each couple of $m_k$ determine a black hole, the $n=4$ solitons solution describes a binary system on Minkowski. 

In that case we remain with a 7-parameter solution: the distance $|z_2-z_1|$ and $m_1, m_2, a_1, a_2, \ell_1, \ell_2$. In the limit where the two sources are very far these parameters can be identified exactly as the mass, angular momentum and NUT charge of each black hole, but if the separation is not large, the back-reaction mixes the physical nature of these constants. 

When $a_2 = \ell_2 = 0$ we have a Kerr-NUT on the side of a Schwarzschild black hole. Indeed taking the $w_4 \to \infty$ limit the event horizon of the static black hole grows indefinitely until it become precisely a Rindler horizon. In that case the resulting metric coincides with the one built with only two solitions on a Rindler accelerating background, which describes an accelerating Kerr-NUT. Remarkably these resulting spacetimes are of type D, therefore they might be included into the Plebanski-Demianski family of black holes. The Petrov algebraic type is computed as in \cite{Podolsky-nut} and \cite{Type-I}.\\
Generically, starting from the $n$-solitonic metric (\ref{metric-ist})-(\ref{metric-ph}), i.e. a series of $n/2$ black holes, $n/2-1$ collinear accelerating Kerr-NUT black holes can be similarly built.\\
On the other hand when $a_2 \neq 0$ or $\ell_2 \neq 0$, the limiting metric is not of D-type; therefore it describes more general accelerating black holes of type-I, such as the ones in the above sections or in \cite{Podolsky-nut} - \cite{PD-NUTs}, as subcases.\\

\subsection{Type D accelerating Kerr-NUT black holes}
\label{sec:kerr-nut}

According to the interpretation of C-metrics as limit of a binary system constituted by a black hole near the horizon of a huge one, we have learned that when the big black hole carries some charges, such as the NUT charge in the previous section or the electric charge as in \cite{Type-I}, the spacetime is of type I. On the contrary when the big black hole is just a Schwarzschild black hole we have a type D solution, as in section \ref{sec:introduction}.\\
Here we show that a particular subclass of the general solution belongs to the Petrov type D, thus it has an overlap with the vacuum version of the Plebanski-Demainski family. In particular we are interested in accelerating, rotating black holes with gravitomagnetic mass. This special class can be obtained from a general binary black hole system where the big black hole is not endowed with angular momentum nor NUT parameter. Therefore, in line with the above parametrisation (\ref{C01}), it means that $a_2=0$ and $\ell_2=0$. Then, after the $w_4 \to \infty$ limit, we obtain exactly the two soliton metric on a Rindler background, which can be defined by
\beq \label{rindler-back}
             \overset{\circ}{g}(\r,z) =  \bigg( \begin{tabular}{cc}
        $-\m_3$ & 0 \\ 
        0 & $\rho^2/\m_3$ 
        \end{tabular} \bigg)    \hspace{3mm} ,  \hspace{1cm} \overset{\circ}{f}(\rho,z) = \frac{\m_3}{\r^2 + \m_3^2} \ \ .
\eeq  
The full metric is given by (\ref{metric-ist})-(\ref{f-ph}) with $n=2$ on the background (\ref{rindler-back}). The integrating constants can be chosen as in (\ref{C01}):
\beq
          C_0^{(2)} = \frac{a+\ell}{2 C_0^{(1)}} \ , \hspace{1cm}  C_1^{(1)} = \frac{-m+\sqrt{m^2+\ell^2-a^2}}{a+\ell} C_0^{(1)} \ , \hspace{1cm} C_1^{(2)} = -\frac{m+\sqrt{m^2+\ell^2-a^2}}{2 C_0^{(1)}}  \ , \nn
\eeq
where $C_0^{(1)}$ can be fixed to 1 without losing generality. This is a suitable parametrization for the non accelerating case, where $m$ corresponds precisely to the mass, $a$ with the angular momentum for unit mass and $b$ with the NUT parameter. Hence, with these values, the metric has all the desired limits in the non-accelerating case (i.e. $w_3 \to \infty$), without angular momentum or NUT. This observation opens to the possibility of having an accelerating Taub-NUT black hole of type D. In practice is sufficient to constraint the angular momentum of the spacetime to be zero, so that the only remaining rotational parameter is related to the NUT charge. \\
However in the presence of acceleration the interpretation of the parameters may vary. For instance, in order to avoid Misner strings, the $\omega(\r,z):=-g_{t\varphi}/g_{tt}$ function on the symmetry axis has to be null. It is possible to constraint one of the parameters $C_a^{(k)}$ by requiring that on the two sectors $(-\infty,w_1)$ and $(w_2,w_3)$ the value of $\omega(\r,z)$ coincides: 
\bea \label{Dw}
           \D\om &=&\lim_{\r \to 0} \Bigg(- \frac{g_{t\varphi}}{g_{tt}} \bigg|_{(-\infty,w_1]} \bigg) - \lim_{\r \to 0} \bigg( -\frac{g_{t\varphi}}{g_{tt}} \bigg|_{[w_2, w_3]} \bigg)   \\
           &=& (w_2-w_1) \left[ \frac{ C_1^{1}C_1^{2}}{(w_3-w_2)C_0^{2}-C_1^{2}} - \frac{C_0^{2}}{(w_3-w_1)C_0^{2} C_1^{1} - (w_3-w_2)C_1^{2}}  \right]= 0\ , \nn
\eea
and then fix it to zero by tuning the arbitrariness of an addictive constant in  $\omega(\r,z)$, which changes the angular velocity of the observer at spatial infinity. A better insight for this spacetime might be achieved in spherical-like coordinates, see section \ref{sec:TypeDacceleratingTaub-NUT}, in particular equations (\ref{KNN-inizio})-(\ref{KNN-fine}).  \\

\subsection{Type D accelerating Taub-NUT}
\label{sec:TypeDacceleratingTaub-NUT}

The accelerating Kerr-NUT metric, written in Weyl coordinates in the section above, take a more intuitive form in a slightly variation of the prolate spheroidal coordinates $(r,x)$, where $r$ can be considered a radial coordinate and $x$ is related with the polar angle $\theta$ through $x:=\cos \theta$. The actual coordinates transformation is given by
\beq
                \rho(r,x):=\frac{\sqrt{\D_r(r) \ \D_x(x)}}{(1+\a rx)^2} \ \ , \hspace{1.5cm} z(r,x):= \frac{(\a r + x)[(r-m)(1+\a m x) + \a x \s^2]}{(1+\a r x)^2} \ , \nn
\eeq  
where
\beq
        \D_r(r):= (1-\a^2 r^2) [(r-m)^2-\s^2] \ \ , \hspace{1.5cm} \D_x(x):=(1-x^2)[(1+\a m x)^2-\a^2x^2\s^2] \ . \nn 
\eeq
Choosing 
\beq
        w_1=-\s \ , \hspace{1.3cm} w_2=\s \ , \hspace{1.3cm} w_3= \frac{1-\a^2(m^2-\s^2)}{2\a} \ ,  \hspace{1.3cm} C_f = - \frac{[1-\a^2(\ell^2-a^2)]^2}{(1+\a^2 a^2) \a^6} \ , \nn
\eeq
\bea
        C_0^{(1)} &=& 1 \ , \hspace{4.9cm}  C_0^{(2)} \ = \ \frac{(a+\ell)[ (1 + \a m)^2 - \a^2\s^2 ]}{2[1+\a^2(m^2-\s^2)]} \ , \nn \\  
         C_1^{(1)} &=& \frac{m-\a m^2 + \s +\a \s^2}{(a+\ell)[1 + \a (m + \s)]} \ , \hspace{1.8cm} C_1^{(2)} \ = \ -\frac{\s}{2} +m\left(\frac{1}{1+\a^2(m^2-\s^2)} - \mezzo \right)  \ , \nn
\eea
the single type D rotating and accelerating black hole endowed with NUT parameter can finally be written as\footnote{A Mathematica notebook with the solution and its fundamental limiting spacetimes can be found between the arXiv source files, for the reader convenience.}
\beq \label{KNN-inizio}
        ds^2 = -f(r,x) \left[ dt - \omega(r,x) d\varphi \right]^2 + \frac{1}{f(r,x)} \left[ e^{2\gamma(r,x)}  \left( \frac{{d r}^2}{\Delta_r(r)} + \frac{{d x}^2}{\Delta_x(x)} \right) + \rho^2(r,x) d\varphi^2 \right] \ \ ,
\eeq
with 
\bea
    f&=&  \frac{(1+\a r x)^{-2} \{[1+\a^2(\ell^2-a^2)x^2]^2\D_r-[a+2\a \ell r + a \a^2 r^2]^2\D_x \}}{\{ \a^2 \ell^4 x^2 + 2 a \ell (\a r  - x) (1-\a rx)+ (1+\a^2 a^2)(r^2+a^2x^2)+\ell^2[1+\a x (\a x(r^2-2a^2)-4r)] \}} \ ,  \nn  \\
    \omega &=& \frac{(a-2\ell x + ax^2)[1+\a^2(\ell^2-a^2)x^2]\D_r + (r^2+\ell^2-a^2)(a+2\a \ell r + \a^2 a r^2 )\D_x}{[1+\a^2(\ell^2-a^2)x^2]^2\D_r-[a+2\a \ell r + a \a^2 r^2]^2\D_x} + \omega_0 \ , \nn \\
  \label{KNN-fine}   \gamma &=& \mezzo \log \left\{\frac{[1+\a^2 x^2(\ell^2-a^2)]^2\D_r-(a+ 2 \a \ell r + a \a^2 r^2)^2\D_x}{(1+\a^2 a^2) (1+\a r x)^4} \right\}   \ .  
\eea
This form of the metric is a significant improvement with respect the previous known parametrizations derived from the Plebabski-Demianski metric \cite{Podolsky-2021}, because it allows to get clear limits to all the possible physical subcases, notably including the elusive type D accelerating Taub-NUT spacetime, which was previously unknown in the literature. In fact, for vanishing the angular momentum parameter, i.e. for $a=0$, we remain with the C-metric-NUT, described by (\ref{KNN-inizio}) with
\bea
    f &=&  \frac{(1+\a^2 \ell^2 x^2)^2\D_r-4\a^2 \ell^2 r^2\D_x}{(1+\a r x)^2 \{r^2 + \ell^2[1+\a x (\a x (\ell^2+r^2)-4r)]\}} \ ,  \nn  \\
 \omega &=&   \frac{2\ell[\a r(r^2+ \ell^2)\D_x - x(1+\a^2\ell^2x^2)\D_r]}{(1+\a^2 \ell^2 x^2)^2\D_r-4\a^2 \ell^2 r^2\D_x}\ +\omega_0 \ ,  \\
  \label{Acc-Taub-NUT}  \gamma &=& \mezzo \log \left[\frac{(1+\a^2 \ell^2 x^2)^2\D_r-4\a^2 \ell^2 r^2\D_x}{(1+\a r x)^4} \right]   \ .  \nn
\eea
This spacetime has not to be confused with the type-I accelerating Taub-NUT metric of section \ref{sec:introduction}. The main difference, as detailed in the previous section, consists in the fact that the background of the type-D is exactly the Rindler spacetime, while the background of the  type-I solution has an extra NUT parameter\footnote{For more details about the accelerating NUT background see \cite{PD-NUTs}.}. \\
Clearly the standard C-metric (\ref{c-metric}) can be obtained from (\ref{KNN-inizio})-(\ref{KNN-fine}) when both the NUT and angular momentum parameters vanish: $\ell \to 0$ and $a \to 0$. While the Taub-NUT spacetime can be reached by vanishing the acceleration and the angular momentum, that is for $\a=0$ and $a=0$\footnote{When the acceleration parameter is null we recover the Kerr-NUT metric of section \ref{sec:kerr-nut}, up to an  sign in an integrating constant.}.\\
On the other hand, from the general solution endowed also with angular momentum (\ref{KNN-inizio})-(\ref{KNN-fine}), we can recover the accelerating Kerr black hole, as written in \cite{hong-teo-rotating-C}, up to an adjustment of the frame of reference described by a trivial change of ($t,\varphi$) coordinates, just by setting $\ell$ to zero.\\
All the limits are well defined and, when the various parameters are switched off, the limiting ordering commute, they can be derived easily from the general metric (\ref{KNN-inizio})-(\ref{KNN-fine}), vanishing one or more integrating constants. This is a non-trivial fact since in the accelerating Kerr-NUT metrics known so far in the literature \cite{Podolsky-2021}, this was not the case. For instance turning off the angular momentum for the Plebanski-Demianski metrics previously known in the literature, would also turn off the acceleration, but the vice-versa was not true. That was the reason why it was not possible to obtain the accelerating-Taub-NUT spacetime from previous accelerating Kerr-NUT solutions. \\
According to the chosen parameterization it is clear that the Misner string is due to the presence of the NUT parameter $\ell$. In fact, from eq. (\ref{Dw}) the discontinuity of the $\omega$ function on the z-axis is
\beq
        \D\omega := \lim_{x \to 1} \left(-\frac{g_{t\varphi}}{g_{tt}} \right) \ - \ \lim_{x \to -1} \left(-\frac{g_{t\varphi}}{g_{tt}} \right) \ =  \  - \frac{4 \ell}{1 + \a^2 (\ell^2-a^2)} \ ,  
\eeq 
thus it cannot be removed when $\ell \neq 0$.\\

\subsection{Type D accelerating Kerr-NUT and accelerating Taub-NUT with cosmological constant}

The generalisation of the accelerating Kerr-NUT spacetime (\ref{KNN-inizio})-(\ref{KNN-fine}) to the presence of the cosmological constant cannot be achieved with the tools provided by  solution generating techniques, because some symmetries determinant for the integrability of the model are broken \cite{melvin-lambda}. Nevertheless thanks to an educated ansatz based on the above solution, the Einstein field equations can be directly integrated; thus it is possible to extend the type-D accelerating Taub-NUT and Kerr-NUT spacetime to the presence of the cosmological constant $\L$. Note that this type-D accelerating Kerr-NUT metric seems not physically equivalent to the standard Plebanski-Demianski, because this latter has no limit to the type D accelerating Taub-NUT. However it is unclear, at the moment, if these two metrics might be diffeomorphic. \\ 
In practice the presence of the cosmological constant doesn't affect the metric form (\ref{KNN-inizio})-(\ref{KNN-fine}), but only only modifies the function $\Delta_r(r)$ and $\D_x(x)$ as follows
\bea
\hspace{-0.6cm}     &&     \D_r(r)  \to  \D_r(r) + \L \left\{ \ell^2 \k + \ell \e r + \left[ a (\a\e +\x) + \a^2 \ell^4 \c - (1+a^2\a^2)(\k-a^2\chi) - \ell^2 \left(\frac{1}{1+a^2\a^2} + 2a^2\a^2\c  \right) \right] r^2 \right. \nn \\
\hspace{-0.6cm}        &&  \hspace{3.3cm} \left. - \ \a \ell r^3\left( \frac{4a}{3+3a^2\a^2} -\x \right) - \left( \frac{1}{3} +\a^2\ell^2 \c \right) r^4  \right\} , \nn \\
\hspace{-0.6cm} &&         \D_x(x)  \to  \D_x(x) + \L \left\{ \ell^2 \c + \ell \x x  + \left(\frac{4a\ell}{3+3a^2 \a^2} +\a \ell \e \right)x^3 - \left( \frac{a^2+(a^2-\ell^2)^2\a^2}{3+3a^2\a^2}  + \ell^2\a^2\k\right) x^4 \right.  \nn  \\
\hspace{-0.6cm} &&   \hspace{3.2cm} \left. + \left[\k+a^2\a^2\k-\a^2\ell^4\c + \ell^2\left( 2a^2\a^2 \c - \frac{1}{1+a^2\a^2} \right) -a \left(\x+a\c+\a(\e+a^3\a\c) \right) \right] x^2 \right\}  , \nn
\eea
where $ \c, \x, \k, \e  $ are integrating constants that can be properly set to simplify the metric  and to obtain the limit to spacetimes without acceleration, angular momentum or without the Misner string. The metric remains of type D according to the Petrov classification.\\
We note that for $a=0$ we get the (type-D) accelerating Taub-NUT metric with cosmological constant, a spacetime previously unknown in the literature. Thus this metric describes the most general type D black hole family in general relativity. \\
Unfortunately the interpretation of the accelerating C-metrics with cosmological constant as a black hole in the near horizon region of another huge black hole cannot be addressed with the theoretical tools at our disposal. In fact, when the cosmological constant is not null, binary black hole systems, such as the ones used here cannot be built, because of the lack of Weyl metrics and solution generating techniques.\\
For further generalisations, in particular in the presence of a Maxwell electromagnetic coupling can be found in \cite{general-D}.\\

\section{Summary, Discussion and Conclusions}

In this article we give a novel interpretation to C-metrics and the Plebanski-Demianski family. Thanks to the Einstein equivalence principle, we interpret the acceleration typical of these accelerating black holes as caused by the presence of another indefinitely large black hole located at a finite distance. In this alternative dual picture the role of external gravitational fields\footnote{While in this article we deal only with gravitational fields to accelerate the back hole, as Ernst have shown in \cite{ernst-remove}, also external electromagnetic fields can be used for the same purpose, however the c-metric must contain an electromagnetic monopole charge.}, or of the cosmic string, is to sustain the black hole at constant distance from the accelerating horizon, instead of providing acceleration by pulling the black hole (as in the usual C-metrics interpretation). So we can interpret the physical system as a black hole near the horizon of another big black hole, whose event horizon becomes an accelerating horizon. \\
In fact we can prove analytically that accelerating black holes can be derived as limits of binary systems, where one of the black hole event horizons grows indefinitely. In particular accelerating type-D black holes, such as C-metric and the Plebanski-Demianski, are derived from a black hole couple where the bigger companion is Schwarzschild-like, that is it carries no extra charges apart from mass. Actually to get the uncharged C-metric both black holes are of Schwarzschild type. When the smaller black hole carries angular momentum, NUT or other charges we retrieve more general accelerating black holes with respect to the Plebanski-Demianski class, but still of type D. Remarkably all the limits to the subcases of the accelerating Kerr-NUT metric are clear and well defined, therefore we are able for the first time to write the metric for the type D accelerating Taub-NUT spacetime, and also its cosmological generalisation. \\
On the other hand when the bigger black hole carries extra properties apart from the mass, these features remain present in the spacetime also after taking the big horizon limit. These extra physical attributes cause the spacetime of being algebraically more general, i.e. of type I. In practice we are left with some residual frame dragging or electromagnetic field related to the rotational parameters or electromagnetic charges \cite{Type-I} reminiscent of the bigger black hole before the large event horizon limit. \\
This interpretation allows us to build, thanks to the inverse scattering techniques, more accelerating general black hole configurations starting from a collinear series of $n$ Kerr-NUT black holes and taking the limit of large event horizon for one of the two black holes in the external position. In this way we get a general $n-1$ series of accelerating collinear Kerr-NUT black holes. Similar considerations for the electromagnetic case can be done with the help of the charged inverse scattering technique \cite{belinski-book}, to obtain a general series of collinear, accelerating Kerr-Newmann black holes.  \\
This picture helps to understand or contextualize some features of accelerating black holes. For instance this interpretation clarifies why the black hole is accelerating: because the interaction with an adjacent huge black companion. Moreover we understand why the conical singularity cannot be removed from accelerating black holes, unless introducing external fields or matter. That's because a gravitational binary system cannot stay at equilibrium, in fact it tends to collapse and it is plagued by conical singularities too. Also we give an interpretation to the nature of the accelerating horizon, which now is considered as the event horizon of an infinitely big black hole close to the black hole. So basically these generalised C-metrics can be considered as the near-horizon metric of a black hole with another small companion nearby.\\

\section*{Acknowledgements}
{\small %We thank Adriano Viganò for interesting discussions on the subject. %M.A. would like to thank the Universidad Austral de Chile (UACh) and the Centro de Estudios Cientificos (CECs) for the hospitality while part of this work was done. 
We thank Ernesto Frodden for useful comments and Jiri Podolsky for interesting discussions on the subject and for the warm hospitality at the Charles University. We thank S.Q. Wu for spotting out several typos.
A Mathematica notebook containing the main solutions presented in this article can be found in the arXiv source folder. %This work has been partially funded by INFN and by MIUR-PRIN contract 2017CC72MK-003} %n$^\textrm{o}$
}

%\appendix

%\section{App 1}
%\label{app:1}


\begin{thebibliography}{99}


\bibitem{Plebanski-Demianski}
  J.~F.~Plebanski and M.~Demianski,
  {\it ``Rotating, charged, and uniformly accelerating mass in general relativity''},
   \href{https://doi.org/10.1016/0003-4916(76)90240-2}{Annals Phys. \textbf{98} (1976), 98-127}.
    %509 citations counted in INSPIRE as of 18 Jan 2023


\bibitem{Podolsky-nut}
J.~Podolsky and A.~Vratny,
{\it ``Accelerating NUT black holes''}, 
  \href{https://doi.org/10.1103/PhysRevD.102.084024}{Phys. Rev. D \textbf{102} (2020) no.8, 084024}; 
   \href{https://arxiv.org/pdf/2007.09169.pdf}{\tt [arXiv:2007.09169 [gr-qc]]}. 
    %9 citations counted in INSPIRE as of 18 Jan 2023


\bibitem{PD-NUTs}
  M.~Astorino and G. Boldi,
     {\it ``Plebanski-Demianski goes NUTs (to remove the Misner string)''},
      \href{https://doi.org/10.1007/JHEP08(2023)085}{JHEP \textbf{08} (2023), 085};
       \href{https://arxiv.org/pdf/2305.03744.pdf}{\tt [arXiv:2305.03744 [gr-qc]]}.


\bibitem{Type-I}
M.~Astorino,
 {\it ``Accelerating and Charged Type I Black Holes''},
  \href{https://doi.org/10.1103/PhysRevD.108.124025}{Phys. Rev. D \textbf{108} (2023) no.12, 124025};
   \href{https://arxiv.org/pdf/2307.10534.pdf}{\tt [arXiv:2307.10534 [gr-qc]]}.


\bibitem{mann-stelea-chng}
B.~Chng, R.~B.~Mann and C.~Stelea,
{\it ``Accelerating Taub-NUT and Eguchi-Hanson solitons in four dimensions''},
  \href{https://doi.org/10.1103/PhysRevD.74.084031}{Phys. Rev. D \textbf{74} (2006), 084031};
   \href{https://arxiv.org/pdf/gr-qc/0608092.pdf}{\tt [arXiv:gr-qc/0608092]}
    %12 citations counted in INSPIRE as of 18 Jan 2023


\bibitem{tesi-giova}
  G.~Boldi,
   {\it ``Ehlers transformation and accelerating spacetimes with a gravomagnetic monopole''},
    \href{https://inspirehep.net/literature/2652409}{Università degli Studi di Milano (2022)}


\bibitem{ernst-magnetic} 
  F.~J.~Ernst,
   {\it ``Black holes in a magnetic universe''},
    \href{https://doi.org/10.1063/1.522781}{J.\ Math.\ Phys.\  {\bf 17}, no. 1, 54 (1976).}


\bibitem{ernst-remove}
  F.~J.~Ernst,
   {\it ``Removal of the nodal singularity of the C-metric''}, 
    \href{http://scitation.aip.org/content/aip/journal/jmp/17/4/10.1063/1.522935}{J. Math. Phys. {\bf 17}, 515 (1976)}.


\bibitem{swirling}
 M.~Astorino, R.~Martelli and A.~Vigan\`o,
  {\it ``Black holes in a swirling universe''},
   \href{https://doi.org/10.1103/PhysRevD.106.064014}{Phys. Rev. D \textbf{106} (2022) no.6, 064014};
    \href{https://arxiv.org/pdf/2205.13548}{\tt [arXiv:2205.13548 [gr-qc]]}.
     %0 citations counted in INSPIRE as of 26 Jun 2022


\bibitem{marcoa-remove}
 M.~Astorino,
  {\it ``Removal of conical singularities from rotating C-metrics and dual CFT entropy''}
   \href{https://doi.org/10.1007/JHEP10(2022)074}{JHEP \textbf{10} (2022), 074};
    \href{https://arxiv.org/pdf/2207.14305}{\tt [arXiv:2207.14305 [gr-qc]]}.


\bibitem{emparan-reall}
 R.~Emparan and H.~S.~Reall,
  {``Generalized Weyl solutions''}, 
   \href{https://doi.org/10.1103/PhysRevD.65.084025}{Phys. Rev. D \textbf{65} (2002), 084025}; 
    \href{https://arxiv.org/pdf/hep-th/0110258.pdf}{\tt [arXiv:hep-th/0110258 [hep-th]]}.
     %280 citations counted in INSPIRE as of 07 Oct 2023


\bibitem{bach-weyl}
 R. Bach and H. Weyl,
  {\it ``Neue l{\"o}sungen der einsteinschen gravitationsgleichungen''},
   \href{https://doi.org/10.1007/BF01485284}{Mathematische Zeitschrift 13, 134–145 (1922)}.

\bibitem{Wang:1996sn}
 Y.~c.~Wang,
  {\it ``Vacuum C metric and the metric of two superposed Schwarzschild black holes''},
   \href{https://doi.org/10.1103/PhysRevD.55.7977}{Phys. Rev. D \textbf{55} (1997), 7977-7979}
    %doi:10.1103/PhysRevD.55.7977
     %7 citations counted in INSPIRE as of 25 Oct 2023

\bibitem{Klemm-embedding}
 A.~Gnecchi, K.~Hristov, D.~Klemm, C.~Toldo and O.~Vaughan,
  {\it ``Rotating black holes in 4d gauged supergravity''}
   \href{https://doi.org/10.1007/JHEP01(2014)127}{JHEP \textbf{01} (2014), 127};
    \href{https://arxiv.org/pdf/1311.1795.pdf}{\tt [arXiv:1311.1795 [hep-th]]}.
     %97 citations counted in INSPIRE as of 24 Oct 2023


\bibitem{ernst-generalized-c} 
 F.~J.~Ernst,
  {\it ``Generalized C-metric''},
   \href{https://doi.org/10.1063/1.523896}{J.\ Math.\ Phys.\  {\bf 19}, 1986-1987 (1978).}

   
\bibitem{many-rotating}
 M.~Astorino and A.~Vigan\`o,
  {\it ``Charged and rotating multi-black holes in an external gravitational field''},
   \href{https://doi.org/10.1140/epjc/s10052-021-09693-6}{Eur. Phys. J. C \textbf{81} (2021) no.10, 891}; 
    \href{https://arxiv.org/abs/2105.02894}{\tt [arXiv:2105.02894 [gr-qc]]}.
     %10 citations counted in INSPIRE as of 17 Jun 2023


\bibitem{marcoa-binary}
 M.~Astorino and A.~Vigan\`o,
  {\it ``Binary black hole system at equilibrium''},
    \href{https://doi.org/10.1016/j.physletb.2021.136506}{Phys. Lett. B \textbf{820} (2021), 136506};
     %doi:10.1016/j.physletb.2021.136506
      \href{https://arxiv.org/pdf/2104.07686}{\tt [2104.07686 [gr-qc]]}.
       %4 citations counted in INSPIRE as of 13 Mar 2022


\bibitem{deCastro}
 G.~M.~de Castro and P.~S.~Letelier,
  {\it ``Black holes surrounded by thin rings and the stability of circular orbits''},
   \href{https://doi.org/10.1088/0264-9381/28/22/225020}{Class. Quant. Grav. \textbf{28} (2011), 225020}


\bibitem{ernst2}
 F.~J.~Ernst,
  {\it ``New Formulation of the Axially Symmetric Gravitational Field Problem. II''},
    \href{https://doi.org/10.1103/PhysRev.168.1415}{\tt Phys.\ Rev.\  {\bf 168} (1968) 1415.}


\bibitem{enhanced}
  M.~Astorino,
  {\it ``Enhanced Ehlers Transformation and the Majumdar-Papapetrou-NUT Spacetime''},
  \href{https://doi.org/10.1007/JHEP01(2020)123}{JHEP \textbf{01} (2020), 123}; 
   %doi:10.1007/JHEP01(2020)123
   \href{https://arxiv.org/pdf/1906.08228}{\tt [arXiv:1906.08228 [gr-qc]]}.
    

\bibitem{belinski-book}
 V. Belinski, E. Verdaguer, \href{https://doi.org/10.1017/CBO9780511535253}{\it ``Gravitational solitons''}, Cambridge, Cambridge Univ. Press, 2001.



\bibitem{Podolsky-2021}
 J.~Podolsky and A.~Vratny,
  {\it ``New improved form of black holes of type D''},
    \href{https://doi.org/10.1103/PhysRevD.104.084078}{Phys. Rev. D \textbf{104} (2021), 084078};
     %doi:10.1103/PhysRevD.104.084078
      \href{https://arxiv.org/pdf/2108.02239}{\tt [arXiv:2108.02239 [gr-qc]]}.
       %9 citations counted in INSPIRE as of 27 Mar 2023

\bibitem{hong-teo-rotating-C}
 K.~Hong and E.~Teo,
  {\it ``A New form of the rotating C-metric''},
   \href{https://doi.org/10.1088/0264-9381/22/1/007}{Class. Quant. Grav. \textbf{22} (2005), 109-118};
    \href{https://arxiv.org/pdf/gr-qc/0410002.pdf}{\tt [arXiv:gr-qc/0410002 [gr-qc]]}.
     %55 citations counted in INSPIRE as of 07 Jan 2024

\bibitem{melvin-lambda}
M.~Astorino,
{\it ``Charging axisymmetric space-times with cosmological constant''},
\href{https://doi.org/10.1007/JHEP06(2012)086}{JHEP \textbf{06} (2012), 086} ;
\href{https://arxiv.org/pdf/1205.6998.pdf}{\tt [arXiv:1205.6998 [gr-qc]]}.
%35 citations counted in INSPIRE as of 27 Sep 2022

\bibitem{general-D}
M.~Astorino,
{\it ``Most general Type D Black Hole and the Accelerating Reissner-Nordstrom-NUT-(A)dS solution'',}
\href{https://doi.org/10.1103/PhysRevD.110.104054}{Phys. Rev. D \textbf{110} (2024), 104054}; \href{https://arxiv.org/pdf/2404.06551.pdf}{\tt [arXiv:2404.06551 [gr-qc]]}.


%%%%%%%%%%%%%%%%%%%%%%%%%%%%%%%%%%%%%%%%%%%%%%%%%%%%%%%%%%%%%%%%%%%%%%%%%%%%%%%%%%%%%%%%%%%%%%%%%%%%%%%%%%%%%%%%%%%%%%%%








\iffalse





\bibitem{Podolsky-nut}
J.~Podolsky and A.~Vratny,
{\it ``Accelerating NUT black holes''}, 
 \href{https://doi.org/10.1103/PhysRevD.102.084024}{Phys. Rev. D \textbf{102} (2020) no.8, 084024}; 
\href{https://arxiv.org/pdf/2007.09169.pdf}{\tt [arXiv:2007.09169 [gr-qc]]}. 
%9 citations counted in INSPIRE as of 18 Jan 2023


\bibitem{Plebanski-Demianski}
  J.~F.~Plebanski and M.~Demianski,
  {\it ``Rotating, charged, and uniformly accelerating mass in general relativity''},
  \href{https://doi.org/10.1016/0003-4916(76)90240-2}{Annals Phys. \textbf{98} (1976), 98-127}.
  %509 citations counted in INSPIRE as of 18 Jan 2023

\bibitem{tesi-giova}
  G.~Boldi,
    {\it ``Ehlers transformation and accelerating spacetimes with a gravomagnetic monopole''},
     \href{https://inspirehep.net/literature/2652409}{Università degli Studi di Milano (2022)}

\bibitem{PD-NUTs}
  M.~Astorino and G. Boldi,
     {\it ``Plebanski-Demianski goes NUTs (to remove the Misner string)''},
     \href{https://doi.org/10.1007/JHEP08(2023)085}{JHEP \textbf{08} (2023), 085};
 \href{https://arxiv.org/pdf/2305.03744.pdf}{\tt [arXiv:2305.03744 [gr-qc]]}.
  
  
\bibitem{ernst-remove}
  F.~J.~Ernst,
  {\it ``Removal of the nodal singularity of the C-metric''}, 
   \href{http://scitation.aip.org/content/aip/journal/jmp/17/4/10.1063/1.522935}{J. Math. Phys. {\bf 17}, 515 (1976)}.


\bibitem{ernst-generalized-c} 
  F.~J.~Ernst,
  {\it ``Generalized C-metric''},
   \href{https://doi.org/10.1063/1.523896}{J.\ Math.\ Phys.\  {\bf 19}, 1986-1987 (1978).}


%\cite{Astorino:2021rdg}
\bibitem{multipolar-acc}
M.~Astorino and A.~Vigan\`o,
{\it``Many accelerating distorted black holes''},
\href{https://doi.org/10.1140/epjc/s10052-021-09693-6}{Eur. Phys. J. C \textbf{81} (2021) no.10, 891};
%doi:10.1140/epjc/s10052-021-09693-6
\href{https://arxiv.org/pdf/2106.02058.pdf}{\tt [2106.02058 [gr-qc]]}.
%6 citations counted in INSPIRE as of 05 Apr 2022

\bibitem{ernst2}
  F.~J.~Ernst,
  {\it ``New Formulation of the Axially Symmetric Gravitational Field Problem. II''},
  \href{https://doi.org/10.1103/PhysRev.168.1415}{\tt Phys.\ Rev.\  {\bf 168} (1968) 1415.}

\bibitem{ernst-hauser}
I.~Hauser and F.~J.~Ernst,
{\it ``Proof of a generalized Geroch conjecture for the hyperbolic Ernst equation''},
\href{https://doi.org/10.1023/A:1002701301339}{Gen. Rel. Grav. \textbf{33} (2001), 195-293}; 
%doi:10.1023/A:1002701301339
\href{https://arxiv.org/pdf/gr-qc/0002049.pdf}{\tt [arXiv:gr-qc/0002049 [gr-qc]]}.
%15 citations counted in INSPIRE as of 11 Jul 2023

\bibitem{many-rotating}
M.~Astorino and A.~Vigan\`o,
{\it ``Charged and rotating multi-black holes in an external gravitational field''},
\href{https://doi.org/10.1140/epjc/s10052-021-09693-6}{Eur. Phys. J. C \textbf{81} (2021) no.10, 891}, 
\href{https://arxiv.org/abs/2105.02894}{\tt [arXiv:2105.02894 [gr-qc]]}.
%10 citations counted in INSPIRE as of 17 Jun 2023

\bibitem{reina-treves}
  A.~Reina and A. Treves
  {\it ``NUT-like generalization of axisymmetric gravitational fields''} ,  \href{https://doi.org/10.1063/1.522614}{Journal of Mathematical Physics 16, 834 (1975)}.

\bibitem{enhanced}
  M.~Astorino,
  {\it ``Enhanced Ehlers Transformation and the Majumdar-Papapetrou-NUT Spacetime''},
  \href{https://doi.org/10.1007/JHEP01(2020)123}{JHEP \textbf{01} (2020), 123} ; 
   %doi:10.1007/JHEP01(2020)123
   \href{https://arxiv.org/pdf/1906.08228}{\tt [arXiv:1906.08228 [gr-qc]]}.
    %3 citations counted in INSPIRE as of 17 Jan 2022
    
\bibitem{ernst-magnetic} 
  F.~J.~Ernst,
  {\it ``Black holes in a magnetic universe''},
  \href{https://doi.org/10.1063/1.522781}{J.\ Math.\ Phys.\  {\bf 17}, no. 1, 54 (1976).}

\bibitem{marcoa-pair}
  M.~Astorino,
  {\it ``Pair Creation of Rotating Black Holes''},
  \href{https://doi.org/10.1103/PhysRevD.89.044022}{Phys. Rev. D \textbf{89} (2014) no.4, 044022};
  %doi:10.1103/PhysRevD.89.044022
  \href{https://arxiv.org/pdf/1312.1723.pdf}{\tt [arXiv:1312.1723~[gr-qc]]}.
  %9 citations counted in INSPIRE as of 20 May 2022

\bibitem{swirling}
M.~Astorino, R.~Martelli and A.~Vigan\`o,
{\it ``Black holes in a swirling universe''},
\href{https://doi.org/10.1103/PhysRevD.106.064014}{Phys. Rev. D \textbf{106} (2022) no.6, 064014} ;
\href{https://arxiv.org/pdf/2205.13548}{\tt [arXiv:2205.13548 [gr-qc]]}.
%0 citations counted in INSPIRE as of 26 Jun 2022

\bibitem{marcoa-thermo}
M.~Astorino,
{\it ``Thermodynamics of Regular Accelerating Black Holes''},
\href{https://doi.org/10.1103/PhysRevD.95.064007}{Phys. Rev. D \textbf{95} (2017) no.6, 064007}
\href{https://arxiv.org/pdf/1612.04387.pdf}{\tt [arXiv:1612.04387 [gr-qc]]}
%54 citations counted in INSPIRE as of 22 Jun 2023

\bibitem{marcoa-removal}
M.~Astorino,
{\it ``Removal of conical singularities from rotating C-metrics and dual CFT entropy''}
\href{https://doi.org/10.1007/JHEP10(2022)074}{JHEP \textbf{10} (2022), 074};
\href{https://arxiv.org/pdf/2207.14305}{\tt [arXiv:2207.14305 [gr-qc]]}.

\bibitem{Griffiths:2009dfa}
J.~B.~Griffiths and J.~Podolsky,
{\it ``Exact Space-Times in Einstein's General Relativity''}, 
\href{https://doi.org/10.1017/CBO9780511635397}{Cambridge University Press, 2009}
%doi:10.1017/CBO9780511635397
%135 citations counted in INSPIRE as of 02 May 2023

\bibitem{stephani-big-book}
  H.~Stephani, D.~Kramer, M.~A.~H.~MacCallum, C.~Hoenselaers and E.~Herlt,
  {``Exact solutions of Einstein's field equations''}, \ 
  \href{https://doi.org/10.1017/CBO9780511535185}{\tt [doi:10.1017/CBO9780511535185]}

\bibitem{bubble}
M.~Astorino, R.~Emparan and A.~Vigan\`o,
{\it ``Bubbles of nothing in binary black holes and black rings, and viceversa''},
\href{https://doi.org/10.1007/JHEP07(2022)007}{JHEP \textbf{07} (2022), 007} ;
\href{https://arxiv.org/pdf/2204.09690.pdf}{\tt [arXiv:2204.09690 [hep-th]]}.
%0 citations counted in INSPIRE as of 25 Jul 2022

%\cite{Majumdar:1947eu}
\bibitem{majumdar}
S.~D.~Majumdar,
{\it ``A class of exact solutions of Einstein's field equations''},
\href{https://doi.org/10.1103/PhysRev.72.390}{Phys. Rev. \textbf{72} (1947), 390-398}
%doi:10.1103/PhysRev.72.390
%457 citations counted in INSPIRE as of 20 Jul 2023

\bibitem{belinski-book}
V. Belinski, E. Verdaguer, \href{https://doi.org/10.1017/CBO9780511535253}{\it ``Gravitational solitons''}, Cambridge, Cambridge Univ. Press, 2001.


\bibitem{alekseev-belinski-2RN}
G.~A.~Alekseev and V.~A.~Belinski,
{\it ``Superposition of fields of two Reissner - Nordstrom sources''},
%doi:10.1142/9789812834300_0022
\href{https://arxiv.org/abs/0710.2515}{\tt [arXiv:0710.2515 [gr-qc]]}.
%%CITATION = doi:10.1142/9789812834300_0022;%%
%10 citations counted in INSPIRE as of 03 May 2019


\bibitem{manko-2007}
V.~S.~Manko,
{\it ``The Double-Reissner-Nordstrom solution and the interaction force between two spherically symmetric charged particles''},
Phys.\ Rev.\ D {\bf 76} (2007) 124032;\
%doi:10.1103/PhysRevD.76.124032
\href{https://doi.org/10.1103/PhysRevD.76.124032}{\tt [arXiv:0710.2158 [gr-qc]]}.
%%CITATION = doi:10.1103/PhysRevD.76.124032;%%
%26 citations counted in INSPIRE as of 03 May 2019

\bibitem{compere-kerr-cft}
G.~Comp\`ere,
{\it ``The Kerr/CFT correspondence and its extensions''},
\href{https://doi.org/10.1007/s41114-017-0003-2}{Living Rev. Rel. \textbf{15} (2012), 11} ;
%doi:10.1007/s41114-017-0003-2
\href{https://arxiv.org/pdf/1203.3561}{\tt [arXiv:1203.3561 [hep-th]]}.%[arXiv:1203.3561 [hep-th]].
%197 citations counted in INSPIRE as of 23 Jun 2022

\bibitem{lucietti-kunduri}
H.~K.~Kunduri, J.~Lucietti and H.~S.~Reall,
{\it ``Near-horizon symmetries of extremal black holes''},
\href{https://doi.org/10.1088/0264-9381/24/16/012}{Class. Quant. Grav. \textbf{24} (2007), 4169-4190} ;
%doi:10.1088/0264-9381/24/16/012
\href{https://arxiv.org/pdf/0705.4214}{\tt [arXiv:0705.4214 [hep-th]]}.%[arXiv:0705.4214 [hep-th]].
%282 citations counted in INSPIRE as of 23 Jun 2022

\bibitem{strominger-kerr-cft}
M.~Guica, T.~Hartman, W.~Song and A.~Strominger,
{\it ``The Kerr/CFT Correspondence''},
\href{https://doi.org/10.1103/PhysRevD.80.124008}{Phys. Rev. D \textbf{80} (2009), 124008} ;
%doi:10.1103/PhysRevD.80.124008
\href{https://arxiv.org/pdf/0809.4266}{\tt [arXiv:0809.4266 [hep-th]]}.%[arXiv:0809.4266 [hep-th]].
%722 citations counted in INSPIRE as of 23 Jun 2022

\bibitem{strominger-duals}
T.~Hartman, K.~Murata, T.~Nishioka and A.~Strominger,
{\it ``CFT Duals for Extreme Black Holes''},
\href{https://doi.org/10.1088/1126-6708/2009/04/019}{JHEP \textbf{04} (2009), 019} ;
%doi:10.1088/1126-6708/2009/04/019
\href{https://arxiv.org/pdf/0811.4393}{\tt [arXiv:0811.4393 [hep-th]]}.%[arXiv:0811.4393 [hep-th]].
%219 citations counted in INSPIRE as of 23 Jun 2022

\bibitem{acc-cft}
M.~Astorino,
{\it ``CFT Duals for Accelerating Black Holes''}
\href{https://doi.org/10.1016/j.physletb.2016.07.019}{Phys. Lett. B \textbf{760} (2016), 393-405} ;
%doi:10.1016/j.physletb.2016.07.019
\href{https://arxiv.org/pdf/1605.06131}{\tt [arXiv:1605.06131 [hep-th]]}.
%32 citations counted in INSPIRE as of 23 Jun 2022


\bibitem{melvin-lambda}
M.~Astorino,
{\it ``Charging axisymmetric space-times with cosmological constant''},
\href{https://doi.org/10.1007/JHEP06(2012)086}{JHEP \textbf{06} (2012), 086} ;
\href{https://arxiv.org/pdf/1205.6998.pdf}{\tt [arXiv:1205.6998 [gr-qc]]}.
%35 citations counted in INSPIRE as of 27 Sep 2022


\bibitem{adolfo}
J. Barrientos and A.~Cisterna,
{\it ``Ehlers Transformations as a Tool for Constructing Accelerating NUT Black Holes''},
\href{https://doi.org/10.1103/PhysRevD.108.024059}{Phys. Rev. D \textbf{108} (2023) no.2, 024059}; 
\href{https://arxiv.org/pdf/2305.03765.pdf}{\tt[arXiv:2305.03765 [gr-qc]]}.

\bibitem{marcoa-embedding}
M.~Astorino,
{\it ``Embedding hairy black holes in a magnetic universe''},
\href{https://doi.org/10.1103/PhysRevD.87.084029}{Phys. Rev. D \textbf{87} (2013) no.8, 084029} ;
\href{https://arxiv.org/pdf/1301.6794}{\tt [arXiv:1301.6794 [gr-qc]]}.
%21 citations counted in INSPIRE as of 21 May 2022

\bibitem{marcoa-stationary}
M.~Astorino,
{\it ``Stationary axisymmetric spacetimes with a conformally coupled scalar field''},
\href{http://doi.org/10.1103/PhysRevD.91.064066}{Phys. Rev. D \textbf{91} (2015), 064066} ;
\href{https://arxiv.org/pdf/1412.3539}{\tt [arXiv:1412.3539 [gr-qc]]}.

%\cite{Astorino:2021dju}
\bibitem{marcoa-binary}
M.~Astorino and A.~Vigan\`o,
{\it ``Binary black hole system at equilibrium''},
\href{https://doi.org/10.1016/j.physletb.2021.136506}{Phys. Lett. B \textbf{820} (2021), 136506};
%doi:10.1016/j.physletb.2021.136506
\href{https://arxiv.org/pdf/2104.07686}{\tt [2104.07686 [gr-qc]]}.
%4 citations counted in INSPIRE as of 13 Mar 2022




\bibitem{enhanced}
  M.~Astorino,
  {\it ``Enhanced Ehlers Transformation and the Majumdar-Papapetrou-NUT Spacetime''},
  \href{https://doi.org/10.1007/JHEP01(2020)123}{JHEP \textbf{01} (2020), 123} ; 
   %doi:10.1007/JHEP01(2020)123
   \href{https://arxiv.org/pdf/1906.08228}{\tt [arXiv:1906.08228 [gr-qc]]}.
    %3 citations counted in INSPIRE as of 17 Jan 2022

\bibitem{ernst-remove}
  F.~J.~Ernst,
  {\it ``Removal of the nodal singularity of the C-metric''}, 
   \href{http://scitation.aip.org/content/aip/journal/jmp/17/4/10.1063/1.522935}{J. Math. Phys. {\bf 17}, 515 (1976)}.

\bibitem{ernst-generalized-c} 
  F.~J.~Ernst,
  {\it ``Generalized C-metric''},
   \href{https://doi.org/10.1063/1.523896}{J.\ Math.\ Phys.\  {\bf 19}, 1986-1987 (1978).}

\bibitem{marcoa-pair}
  M.~Astorino,
  {\it ``Pair Creation of Rotating Black Holes''},
  \href{https://doi.org/10.1103/PhysRevD.89.044022}{Phys. Rev. D \textbf{89} (2014) no.4, 044022};
  %doi:10.1103/PhysRevD.89.044022
  \href{https://arxiv.org/pdf/1312.1723.pdf}{\tt [arXiv:1312.1723~[gr-qc]]}.
  %9 citations counted in INSPIRE as of 20 May 2022

\bibitem{multipolar-acc}
  M.~Astorino and A.~Vigan\`o,
  {\it``Many accelerating distorted black holes''},
  \href{https://doi.org/10.1140/epjc/s10052-021-09693-6}{Eur. Phys. J. C \textbf{81} (2021) no.10, 891};
  %doi:10.1140/epjc/s10052-021-09693-6
  \href{https://arxiv.org/pdf/2106.02058.pdf}{\tt [2106.02058 [gr-qc]]}.
  %6 citations counted in INSPIRE as of 05 Apr 2022

\bibitem{misner-counterexample}
  C. W. Misner,  
  {\it ``Taub-NUT space as a counterexample to almost anything''}, 
  Relativity theory and astrophysics 1 (1967): 160.\\  \href{https://ntrs.nasa.gov/api/citations/19660007407/downloads/19660007407.pdf}{\tt [https://ntrs.nasa.gov/api/citations/19660007407/downloads/19660007407.pdf]}

\bibitem{Plebanski-Demianski}
  J.~F.~Plebanski and M.~Demianski,
  {\it ``Rotating, charged, and uniformly accelerating mass in general relativity''},
  \href{https://doi.org/10.1016/0003-4916(76)90240-2}{Annals Phys. \textbf{98} (1976), 98-127}.
  %509 citations counted in INSPIRE as of 18 Jan 2023

\bibitem{Podolsky-2020}
J.~Podolsky and A.~Vratny,
{\it ``Accelerating NUT black holes''}, 
 \href{https://doi.org/10.1103/PhysRevD.102.084024}{Phys. Rev. D \textbf{102} (2020) no.8, 084024}; 
\href{https://arxiv.org/pdf/2007.09169.pdf}{\tt [arXiv:2007.09169 [gr-qc]]}. 
%9 citations counted in INSPIRE as of 18 Jan 2023

\bibitem{mann-stelea-chng}
B.~Chng, R.~B.~Mann and C.~Stelea,
{\it ``Accelerating Taub-NUT and Eguchi-Hanson solitons in four dimensions''},
 \href{https://doi.org/10.1103/PhysRevD.74.084031}{Phys. Rev. D \textbf{74} (2006), 084031} ;
\href{https://arxiv.org/pdf/gr-qc/0608092.pdf}{\tt [arXiv:gr-qc/0608092]}
%12 citations counted in INSPIRE as of 18 Jan 2023

\bibitem{reina-treves}
  A.~Reina and A. Treves
  {\it ``NUT-like generalization of axisymmetric gravitational fields''} ,  \href{https://doi.org/10.1063/1.522614}{Journal of Mathematical Physics 16, 834 (1975)}.

\bibitem{swirling}
M.~Astorino, R.~Martelli and A.~Vigan\`o,
{\it ``Black holes in a swirling universe''},
\href{https://doi.org/10.1103/PhysRevD.106.064014}{Phys. Rev. D \textbf{106} (2022) no.6, 064014} ;
\href{https://arxiv.org/pdf/2205.13548}{\tt [arXiv:2205.13548 [gr-qc]]}.
%0 citations counted in INSPIRE as of 26 Jun 2022

\bibitem{marcoa-removal}
M.~Astorino,
{\it ``Removal of conical singularities from rotating C-metrics and dual CFT entropy''}
\href{https://doi.org/10.1007/JHEP10(2022)074}{JHEP \textbf{10} (2022), 074};
\href{https://arxiv.org/pdf/2207.14305}{\tt [arXiv:2207.14305 [gr-qc]]}.


\bibitem{Podolsky-2021}
J.~Podolsky and A.~Vratny,
{\it ``New improved form of black holes of type D''},
\href{https://doi.org/10.1103/PhysRevD.104.084078}{Phys. Rev. D \textbf{104} (2021), 084078}
%doi:10.1103/PhysRevD.104.084078
\href{https://arxiv.org/pdf/2108.02239}{\tt [arXiv:2108.02239 [gr-qc]]}.
%9 citations counted in INSPIRE as of 27 Mar 2023

\bibitem{Podolsky-2022}
J.~Podolsky and A.~Vratny,
{\it ``New form of all black holes of type D with a cosmological constant''}, 
\href{https://doi.org/10.1103/PhysRevD.107.084034}{Phys. Rev. D \textbf{107} (2023) no.8, 084034}
%doi:10.1103/PhysRevD.107.084034
\href{https://arxiv.org/pdf/2212.08865}{\tt [arXiv:2212.08865 [gr-qc]]}.
%1 citations counted in INSPIRE as of 02 May 2023


\bibitem{bonnor}
W.~B.~Bonnor,
{\it ``A new interpretation of the NUT metric in general relativity''},
\href{https://doi.org/10.1017/s0305004100044807}{Math. Proc. Cambridge Phil. Soc. \textbf{66} (1969) no.1, 145-151}
%42 citations counted in INSPIRE as of 02 May 2023

\bibitem{belinski-book}
V. Belinski, E. Verdaguer, \href{https://doi.org/10.1017/CBO9780511535253}{\it ``Gravitational solitons''}, Cambridge, Cambridge Univ. Press, 2001.

\bibitem{marcoa-embedding}
M.~Astorino,
{\it ``Embedding hairy black holes in a magnetic universe''},
\href{https://doi.org/10.1103/PhysRevD.87.084029}{Phys. Rev. D \textbf{87} (2013) no.8, 084029} ;
\href{https://arxiv.org/pdf/1301.6794}{\tt [arXiv:1301.6794 [gr-qc]]}.
%21 citations counted in INSPIRE as of 21 May 2022

\bibitem{marcoa-stationary}
M.~Astorino,
{\it ``Stationary axisymmetric spacetimes with a conformally coupled scalar field''},
\href{http://doi.org/10.1103/PhysRevD.91.064066}{Phys. Rev. D \textbf{91} (2015), 064066} ;
\href{https://arxiv.org/pdf/1412.3539}{\tt [arXiv:1412.3539 [gr-qc]]}.
%20 citations counted in INSPIRE as of 21 May 2022

\bibitem{alekseev-belinski-kerr}
  G.~A.~Alekseev and V.~A.~Belinski,
  {\it ``Superposition of fields of two rotating charged masses in general relativity and existence of equilibrium configurations''},
  Gen.\ Rel.\ Grav.\  {\bf 51} (2019) no.5,  68
  %doi:10.1007/s10714-019-2543-0
  \href{https://arxiv.org/abs/1905.05317}{\tt [arXiv:1905.05317 [gr-qc]]}.
  %%CITATION = doi:10.1007/s10714-019-2543-0;%%
  %1 citations counted in INSPIRE as of 22 Oct 2019

\bibitem{bubble}
M.~Astorino, R.~Emparan and A.~Vigan\`o,
{\it ``Bubbles of nothing in binary black holes and black rings, and viceversa''},
\href{https://doi.org/10.1007/JHEP07(2022)007}{JHEP \textbf{07} (2022), 007} ;
\href{https://arxiv.org/pdf/2204.09690.pdf}{\tt [arXiv:2204.09690 [hep-th]]}.
%0 citations counted in INSPIRE as of 25 Jul 2022

\bibitem{tesi-giova}
G.~Boldi,
{\it ``Ehlers transformation and accelerating spacetimes with a gravomagnetic monopole''},
\href{https://inspirehep.net/literature/2652409}{\tt Università degli Studi di Milano (2022)}

\bibitem{asto-lambda}
M.~Astorino,
{\it ``Charging axisymmetric space-times with cosmological constant''},
\href{https://doi.org/10.1007/JHEP06(2012)086}{JHEP \textbf{06} (2012), 086} ;
\href{https://arxiv.org/pdf/1205.6998.pdf}{\tt [arXiv:1205.6998 [gr-qc]]}.
%35 citations counted in INSPIRE as of 27 Sep 2022

\bibitem{adolfo}
J. Barrientos and A.~Cisterna,
{\it ``Ehlers Transformations as a Tool for Constructing Accelerating NUT Black Holes''},
\href{https://arxiv.org/pdf/2305.03765.pdf}{\tt[arXiv:2305.03765 [gr-qc]]}.
%0 citations counted in INSPIRE as of 16 May 2023{\it ``Ehlers Transformations as a Tool for Constructing Accelerating NUT Black Holes''}, to appear soon.


\bibitem{ernst-remove}
  F. Ernst
  {\it ``Removal of the nodal singularity of the C-metric''}, 
   \href{http://scitation.aip.org/content/aip/journal/jmp/17/4/10.1063/1.522935}{J. Math. Phys. {\bf 17}, 515 (1976)}.


\bibitem{marcoa-pair}
M.~Astorino,
{\it ``Pair Creation of Rotating Black Holes''},
\href{https://doi.org/10.1103/PhysRevD.89.044022}{Phys. Rev. D \textbf{89} (2014) no.4, 044022};
%doi:10.1103/PhysRevD.89.044022
\href{https://arxiv.org/pdf/1312.1723.pdf}{\tt [arXiv:1312.1723~[gr-qc]]}.
%9 citations counted in INSPIRE as of 20 May 2022

\bibitem{ernst-magnetic} 
  F.~J.~Ernst,
  {\it ``Black holes in a magnetic universe''},
  \href{https://doi.org/10.1063/1.522781}{J.\ Math.\ Phys.\  {\bf 17}, no. 1, 54 (1976).}


\bibitem{ernst-generalized-c} 
  F.~J.~Ernst,
  {\it ``Generalized C-metric''},
   \href{https://doi.org/10.1063/1.523896}{J.\ Math.\ Phys.\  {\bf 19}, 1986-1987 (1978).}


%\cite{Astorino:2021dju}
\bibitem{marcoa-binary}
M.~Astorino and A.~Vigan\`o,
{\it ``Binary black hole system at equilibrium''},
\href{https://doi.org/10.1016/j.physletb.2021.136506}{Phys. Lett. B \textbf{820} (2021), 136506};
%doi:10.1016/j.physletb.2021.136506
\href{https://arxiv.org/pdf/2104.07686}{\tt [2104.07686 [gr-qc]]}.
%4 citations counted in INSPIRE as of 13 Mar 2022


\bibitem{deCastro}
G.~M.~de Castro and P.~S.~Letelier,
{\it ``Black holes surrounded by thin rings and the stability of circular orbits''},
\href{https://doi.org/10.1088/0264-9381/28/22/225020}{Class. Quant. Grav. \textbf{28} (2011), 225020}

%\cite{Astorino:2021rdg}
\bibitem{multipolar-acc}
M.~Astorino and A.~Vigan\`o,
{\it``Many accelerating distorted black holes''},
\href{https://doi.org/10.1140/epjc/s10052-021-09693-6}{Eur. Phys. J. C \textbf{81} (2021) no.10, 891};
%doi:10.1140/epjc/s10052-021-09693-6
\href{https://arxiv.org/pdf/2106.02058.pdf}{\tt [2106.02058 [gr-qc]]}.
%6 citations counted in INSPIRE as of 05 Apr 2022

\bibitem{swirling}
M.~Astorino, R.~Martelli and A.~Vigan\`o,
{\it ``Black holes in a swirling universe''},
\href{https://arxiv.org/pdf/2205.13548}{\tt [arXiv:2205.13548 [gr-qc]]}.
%0 citations counted in INSPIRE as of 26 Jun 2022

\bibitem{reina-treves}
  A.~Reina and A. Treves
  {\it ``NUT-like generalization of axisymmetric gravitational fields''},
  \href{https://doi.org/10.1063/1.522614}{Journal of Mathematical Physics 16, 834 (1975)}.

\bibitem{enhanced}
M.~Astorino,
{\it ``Enhanced Ehlers Transformation and the Majumdar-Papapetrou-NUT Spacetime''},
\href{https://doi.org/10.1007/JHEP01(2020)123}{JHEP \textbf{01} (2020), 123} ; 
%doi:10.1007/JHEP01(2020)123
\href{https://arxiv.org/pdf/1906.08228}{\tt [arXiv:1906.08228 [gr-qc]]}.
%3 citations counted in INSPIRE as of 17 Jan 2022


\bibitem{marcoa-embedding}
M.~Astorino,
{``Embedding hairy black holes in a magnetic universe''},
\href{https://doi.org/10.1103/PhysRevD.92.104006}{Phys. Rev. D \textbf{92} (2015) no.10, 104006};
\href{https://doi.org/10.1103/PhysRevD.87.084029}{Phys. Rev. D \textbf{87} (2013) no.8, 084029} ;
\href{https://arxiv.org/pdf/1301.6794}{\tt [arXiv:1301.6794 [gr-qc]]}.
%21 citations counted in INSPIRE as of 21 May 2022

\bibitem{marcoa-stationary}
M.~Astorino,
{\it ``Stationary axisymmetric spacetimes with a conformally coupled scalar field''},
\href{http://doi.org/10.1103/PhysRevD.91.064066}{Phys. Rev. D \textbf{91} (2015), 064066} ;
\href{https://arxiv.org/pdf/1412.3539}{\tt [arXiv:1412.3539 [gr-qc]]}.
%20 citations counted in INSPIRE as of 21 May 2022

\bibitem{magnetised-kerr-cft}
M.~Astorino,
{ \it ``Magnetised Kerr/CFT correspondence''},
\href{https://doi.org/10.1016/j.physletb.2015.10.017}{Phys. Lett. B \textbf{751} (2015), 96-106};
\href{https://arxiv.org/pdf/1508.01583}{\tt [arXiv:1508.01583 [hep-th]]}.
%31 citations counted in INSPIRE as of 23 Jun 2022

\bibitem{magnetised-RN-cft}
M.~Astorino,
{\it ``Microscopic Entropy of the Magnetised Extremal Reissner-Nordstrom Black Hole''},
\href{https://doi.org/10.1007/JHEP10(2015)016}{JHEP \textbf{10} (2015), 016} ;
\href{https://arxiv.org/pdf/1507.04347}{\tt [arXiv:1507.04347 [hep-th]]}.
%18 citations counted in INSPIRE as of 23 Jun 2022


\bibitem{c-cft}
M.~Astorino,
{\it ``CFT Duals for Accelerating Black Holes''}
\href{https://doi.org/10.1016/j.physletb.2016.07.019}{Phys. Lett. B \textbf{760} (2016), 393-405} ;
%doi:10.1016/j.physletb.2016.07.019
\href{https://arxiv.org/pdf/1605.06131}{\tt [arXiv:1605.06131 [hep-th]]}.
%32 citations counted in INSPIRE as of 23 Jun 2022


\bibitem{lucietti-kunduri}
H.~K.~Kunduri, J.~Lucietti and H.~S.~Reall,
{\it ``Near-horizon symmetries of extremal black holes''},
\href{https://doi.org/10.1088/0264-9381/24/16/012}{Class. Quant. Grav. \textbf{24} (2007), 4169-4190} ;
%doi:10.1088/0264-9381/24/16/012
\href{https://arxiv.org/pdf/0705.4214}{\tt [arXiv:0705.4214 [hep-th]]}.%[arXiv:0705.4214 [hep-th]].
%282 citations counted in INSPIRE as of 23 Jun 2022

\bibitem{strominger-kerr-cft}
M.~Guica, T.~Hartman, W.~Song and A.~Strominger,
{\it ``The Kerr/CFT Correspondence''},
\href{https://doi.org/10.1103/PhysRevD.80.124008}{Phys. Rev. D \textbf{80} (2009), 124008} ;
%doi:10.1103/PhysRevD.80.124008
\href{https://arxiv.org/pdf/0809.4266}{\tt [arXiv:0809.4266 [hep-th]]}.%[arXiv:0809.4266 [hep-th]].
%722 citations counted in INSPIRE as of 23 Jun 2022

\bibitem{strominger-duals}
T.~Hartman, K.~Murata, T.~Nishioka and A.~Strominger,
{\it ``CFT Duals for Extreme Black Holes''},
\href{https://doi.org/10.1088/1126-6708/2009/04/019}{JHEP \textbf{04} (2009), 019} ;
%doi:10.1088/1126-6708/2009/04/019
\href{https://arxiv.org/pdf/0811.4393}{\tt [arXiv:0811.4393 [hep-th]]}.%[arXiv:0811.4393 [hep-th]].
%219 citations counted in INSPIRE as of 23 Jun 2022

\bibitem{compere-kerr-cft}
G.~Comp\`ere,
{\it ``The Kerr/CFT correspondence and its extensions''},
\href{https://doi.org/10.1007/s41114-017-0003-2}{Living Rev. Rel. \textbf{15} (2012), 11} ;
%doi:10.1007/s41114-017-0003-2
\href{https://arxiv.org/pdf/1203.3561}{\tt [arXiv:1203.3561 [hep-th]]}.%[arXiv:1203.3561 [hep-th]].
%197 citations counted in INSPIRE as of 23 Jun 2022

\bibitem{gaston}
L.~Donnay, G.~Giribet, H.~A.~Gonz\'alez and M.~Pino,
{\it ``Extended Symmetries at the Black Hole Horizon''},
\href{https://doi.org/10.1007/JHEP09(2016)100}{JHEP \textbf{09} (2016), 100} ;
%doi:10.1007/JHEP09(2016)100
\href{https://arxiv.org/pdf/1607.05703}{\tt [arXiv:1607.05703 [hep-th]]}.%[arXiv:1607.05703 [hep-th]].
%132 citations counted in INSPIRE as of 24 Jun 2022

\bibitem{bicak}
J.~Bi\v{c}\'ak and F.~Hejda,
{\it ``Near-horizon description of extremal magnetized stationary black holes and Meissner effect''},
\href{https://doi.org/10.1103/PhysRevD.92.104006}{Phys. Rev. D \textbf{92} (2015) no.10, 104006}; 
\href{https://arxiv.org/pdf/1510.01911.pdf}{\tt [arXiv:1510.01911 [gr-qc]]}.
%28 citations counted in INSPIRE as of 24 Jun 2022

\bibitem{ernst2}
F.~J.~Ernst,
  {\it ``New Formulation of the Axially Symmetric Gravitational Field Problem. II''},
  \href{https://doi.org/10.1103/PhysRev.168.1415}{Phys.\ Rev.\  {\bf 168} (1968) 1415}.

\bibitem{strominger}
D.~Garfinkle, S.~B.~Giddings and A.~Strominger,
{\it ``Entropy in black hole pair production''},
\href{https://doi.org/10.1103/PhysRevD.49.958}{Phys. Rev. D \textbf{49} (1994), 958-965};
%doi:10.1103/PhysRevD.49.958
\href{https://arxiv.org/pdf/gr-qc/9306023.pdf}{\tt [arXiv:gr-qc/9306023]}
%116 citations counted in INSPIRE as of 20 May 2022

\bibitem{hawking}
S.~W.~Hawking, G.~T.~Horowitz and S.~F.~Ross,
{\it ``Entropy, Area, and black hole pairs''},
\href{https://doi.org/10.1103/PhysRevD.51.4302}{Phys. Rev. D \textbf{51} (1995), 4302-4314};
%doi:10.1103/PhysRevD.51.4302
\href{https://arxiv.org/pdf/gr-qc/9409013.pdf}{\tt [arXiv:gr-qc/9409013]}.
%389 citations counted in INSPIRE as of 20 May 2022

\bibitem{KramerNeug}
D.~Kramer, G.~Neugebauer,
{\it ``The superposition of two Kerr solutions''},
\href{https://www.sciencedirect.com/science/article/abs/pii/0375960180905563}{Physics Letters A \textbf{75} (1980), no.4, 259--261}.

\bibitem{bubble}
M.~Astorino, R.~Emparan and A.~Vigan\`o,
{\it ``Bubbles of nothing in binary black holes and black rings, and viceversa''},
\href{https://doi.org/10.1007/JHEP07(2022)007}{JHEP \textbf{07} (2022), 007} ;
\href{https://arxiv.org/pdf/2204.09690.pdf}{\tt [arXiv:2204.09690 [hep-th]]}.
%0 citations counted in INSPIRE as of 25 Jul 2022


\bibitem{asto-lambda}
M.~Astorino,
{\it ``Charging axisymmetric space-times with cosmological constant''},
\href{https://doi.org/10.1007/JHEP06(2012)086}{JHEP \textbf{06} (2012), 086} ;
\href{https://arxiv.org/pdf/1205.6998.pdf}{\tt [arXiv:1205.6998 [gr-qc]]}.
%35 citations counted in INSPIRE as of 27 Sep 2022


\bibitem{kerr-magnetic}  
  F.~J.~Ernst and W. Wild,
  {\it ``Kerr black holes in a magnetic universe''},
   J.\ Math.\ Phys.{\bf 17} (1976) 182.


\bibitem{misner-couterexample}
C. W. Misner,  
{\it Taub-NUT space as a counterexample to almost anything},
Relativity theory and astrophysics 1 (1967): 160.\\
\href{https://ntrs.nasa.gov/api/citations/19660007407/downloads/19660007407.pdf}{\tt [https://ntrs.nasa.gov/api/citations/19660007407/downloads/19660007407.pdf]}


 \bibitem{geroch1}
  R.~P.~Geroch,
  {\it ``A Method for generating solutions of Einstein's equations''},
  J.\ Math.\ Phys.\  {\bf 12} (1971) 918.
  %%CITATION = JMAPA,12,918;%%
  %310 citations counted in INSPIRE as of 27 May 2014
  
  \bibitem{geroch2}
  R.~P.~Geroch,
  {\it ``A Method for generating new solutions of Einstein's equation. 2''},
  J.\ Math.\ Phys.\  {\bf 13} (1972) 394.
  %%CITATION = JMAPA,13,394;%%
  %227 citations counted in INSPIRE as of 27 May 2014

\bibitem{hauser-ernst}
I.~Hauser and F.~J.~Ernst,
{\it ``Proof of a generalized Geroch conjecture for the hyperbolic Ernst equation''},
Gen. Rel. Grav. \textbf{33} (2001), 195-293,
%\href{https://doi.org/10.1023/A:1002701301339}{doi:10.1023/A:1002701301339}
\href{https://arxiv.org/abs/gr-qc/0002049}{\tt[arXiv:gr-qc/0002049]}.
 
 
\bibitem{ernst1}
   F.~J.~Ernst,
  {\it ``New formulation of the axially symmetric gravitational field problem''},
  Phys.\ Rev.\  {\bf 167} (1968) 1175.

\bibitem{ernst2}
F.~J.~Ernst,
  {\it ``New Formulation of the Axially Symmetric Gravitational Field Problem. II''},
  Phys.\ Rev.\  {\bf 168} (1968) 1415. \href{https://doi.org/10.1103/PhysRev.168.1415}{\tt [doi:10.1103/PhysRev.168.1415]}

\bibitem{belinski}
V. Belinski, E. Verdaguer, {\it ``Gravitational solitons''}, Cambridge, Cambridge Univ. Press, 2001.



\bibitem{stephani}
  H.~Stephani, D.~Kramer, M.~A.~H.~MacCallum, C.~Hoenselaers and E.~Herlt,
  {``Exact solutions of Einstein's field equations''}, \ 
  \href{https://doi.org/10.1017/CBO9780511535185}{\tt [doi:10.1017/CBO9780511535185]}


\fi

\end{thebibliography}
\end{document}